\documentclass[12pt,a4paper]{article}
\usepackage{amsmath,amssymb,amsthm}
\usepackage[margin=1.0in]{geometry}
\usepackage{cite}
\usepackage{graphicx}
\usepackage{enumerate}
\allowdisplaybreaks
\usepackage[colorlinks=true
,urlcolor=blue
,anchorcolor=blue
,citecolor=blue
,filecolor=blue
,linkcolor=blue
,menucolor=blue
,pagecolor=blue
,linktocpage=true
,pdfproducer=medialab
,pdfa=true
]{hyperref}
\numberwithin{equation}{section}
\def\a{\alpha}
\def\b{\beta}
\def\g{\gamma}
\def\d{\delta}
\def\e{\epsilon}

\def\th{\theta}

\def\k{\kappa}
\def\l{\lambda}
\def\m{\mu}
\def\n{\nu}

\def\r{\rho}

\def\s{\sigma}

\def\ph{\phi}

\def\D{\Delta}

\def\Th{\Theta}

\def\L{\Lambda}

\def\be{\begin{equation}}
\def\ee{\end{equation}}
\def\bea{\begin{eqnarray}}
\def\eea{\end{eqnarray}}

\def\pa{\partial}

\def\lp{\left(}
\def\rp{\right)}
\def\ls{\left[}
\def\rs{\right]}
\def\nn{\nonumber}
\def\ie{{\it i.e., }}

\makeatletter
\renewcommand\section{\@startsection {section}{1}{\z@}%
	{-3.5ex \@plus -1ex \@minus -.2ex}%nn
	{2.3ex \@plus.2ex}%
	{\normalfont\large\bfseries}}
\renewcommand\subsection{\@startsection{subsection}{2}{\z@}%
	{-3.25ex\@plus -1ex \@minus -.2ex}%
	{1.5ex \@plus .2ex}%
	{\normalfont\bfseries}}
\makeatother

\begin{document}

\begin{center}
\addtolength{\baselineskip}{.5mm}
\thispagestyle{empty}
\begin{flushright}
%\today \\
%{\sc }
\end{flushright}

\vspace{20mm}

{\Large \bf Rotating black strings beyond Maxwell's electrodynamics}
\\[15mm]
{Hamid R. Bakhtiarizadeh${}^{a,}$\footnote{h.bakhtiarizadeh@kgut.ac.ir} and Hanif Golchin${}^{b,}$\footnote{h.golchin@uk.ac.ir}}
\\[5mm]
{\it ${}^a$Department of Nanotechnology, Graduate University of Advanced Technology,\\ Kerman, Iran\\${}^b$Faculty of Physics, Shahid Bahonar University of Kerman, P.O. Box 76175,\\ Kerman, Iran}

\vspace{20mm}

{\bf  Abstract}
\end{center}

In this paper, we investigate the asymptotically Anti–de Sitter solutions of rotating black strings coupled to Born-Infeld and Modified Maxwell non-linear electrodynamics in the context of Einsteinian, Einsteinian cubic and Einsteinian quartic gravity. By studying the near-horizon behavior of solutions, we find the mass parameter, surface gravity and accordingly the Hawking temperature. We also compute the entropy, mass, angular momentum, electric charge, and the electrostatic potential of solutions to show that, in the context of above theories and in the presence of two mentioned non-linear electrodynamics, the first law of thermodynamics for rotating black strings is also exactly hold. We also investigate extremality, thermal stability and phase transition of solutions.
\vfill
\newpage

%\tableofcontents

\section{Introduction}\label{int}

Theories of Non-Linear Electrodynamics (NLE) \cite{Sorokin:2021tge}, that approach Maxwell theory in the weak field limit, have been extensively studied as a plausible extension of Maxwell theory to find new physics. The most notable examples are the Born-Infeld (BI) theory \cite{Born:1934gh}, which first was introduced to ensure that electrical self-energy of a charged point particle is finite but much later turned out to be an important issue in string theory, and the Heisenberg-Euler effective theory of quantum electrodynamics \cite{Heisenberg:1936nmg}. The newest recently found NLE is Modified Maxwell (ModMax) theory, the unique Non-Linear (NL) modification of Maxwell’s theory that preserves all its symmetries including electric-magnetic duality as well as conformal invariance \cite{Bandos:2020jsw,Kosyakov:2020wxv}.

In the following, we briefly review some attempts in finding the black hole solutions in the context of BI and ModMax theories. In the context of BI theory, in \cite{Breton:2003tk}, the BI black hole in the isolated horizon framework is probed. In \cite{Fernando:2003tz}, black hole solutions to Einstein-BI gravity with a cosmological constant is constructed. The asymptotically Anti–de Sitter (AdS) and dS black hole solutions of Einstein-BI theory in arbitrary dimension and their geometries and thermodynamic properties are also discussed in \cite{Dey:2004yt}. Some exact solutions in a $ D \geq 4 $-dimensional Einstein-BI theory with a cosmological constant are studied in \cite{Cai:2004eh}. In \cite{Stefanov:2007qw}, numerical solutions describing charged black holes coupled to BI type NLE in scalar-tensor theories of gravity with massless scalar field are found. Also, in \cite{Gullu:2010pc}, a three-dimensional gravitational BI theory which reduces to the New Massive Gravity (NMG) at the quadratic level in the small curvature expansion and at the cubic order reproduces the deformation of NMG obtained from AdS/CFT is presented. Thermodynamics of BI Black Holes is studied in \cite{Chemissany:2008fy}. A new class of asymptotically (A)dS black hole solutions of Einstein-Yang-Mills massive gravity in the presence of BI NLE is constructed in \cite{Hendi:2018hdo}. Holographic complexity of BI black holes is also investigated in \cite{Meng:2018vtl}. In \cite{Dehghani:2019noj}, thermodynamic properties of the three-dimensional charged dilatonic black holes are considered in the presence of BI NLE. Shadow, deflection angle and quasinormal modes of BI black holes are investigated in \cite{Jafarzade:2020ova}. BI black holes are also considered in the context of $ 4D $ Einstein–Gauss–Bonnet gravity \cite{Yang:2020jno} as well as Einsteinian cubic gravity \cite{KordZangeneh:2020qeg}.

In the context of ModMax theory, in \cite{Bandos:2020hgy}, multiple formulations of generic $ SO(2) $-duality invariant NLE theories in four dimensions and generic nonlinear chiral $ 2 $-form electrodynamics in six dimensions have been explored. A prescription for $ N=1 $ supersymmetrization of any NLE theory in four dimensions with a Lagrangian density satisfying a convexity condition is given in \cite{Bandos:2021rqy}. A new generalized ModMax model of nonlinear electrodynamics with four parameters is also proposed in \cite{Kruglov:2021bhs}. In \cite{Kuzenko:2021qcx}, the general theory of $ U(1) $ duality-invariant NLE is introduced and its $ N = 1,2 $ supersymmetric extensions to the cases of bosonic conformal spin-s gauge fields with $ s \geq 2 $ and their $ N = 1,2 $ superconformal cousins is investigated. A Lagrangian for general NLE that features electric and magnetic potentials on equal footing is also proposed in \cite{Avetisyan:2021heg}. Higher-derivative extensions of Einstein-Maxwell theory that are invariant under electromagnetic duality rotations, allowing for non-minimal couplings between gravity and the gauge field are also investigated in \cite{Cano:2021tfs}. The deformation of the ModMax theory, under $ T\bar{T} $-like flows is investigeted in \cite{Babaei-Aghbolagh:2022uij} and it is shown that the deformed theory is the generalized BI NLE. Also, in \cite{Babaei-Aghbolagh:2022itg}, a manifestly $ SL(2, {\mathbb R}) $-invariant form for the energy-momentum tensor of ModMax theory, is investigated. The asymptotically AdS accelerated black holes in ModMax theory and their thermodynamics is studied in \cite{Barrientos:2022bzm}. Magnetic black holes with generalized ModMax model of NLE are explored in \cite{Kruglov:2022qag}. It also is shown that ModMax theory of NLE and its BI-like generalization are related by a flow equation driven by a quadratic combination of stress-energy tensors \cite{Ferko:2022iru}.

In this paper, we will focus on finding the rotating black string solutions \cite{Lemos:1994xp,Lemos:1995cm} in the context of Einsteinian gravity (EG), Einsteinian Cubic Gravity (ECG) and Einsteinian Quartic Gravity (EQG) with BI as well as ModMax source. The solutions of charged rotating black strings, are generalized to include higher dimensions in \cite{Awad:2002cz} and their thermodynamic properties are also investigated in \cite{Dehghani:2002rr,Dehghani:2002jh}. Solutions are also explored in the presence of BI and power Maxwell invariant \cite{Hendi:2010kv} as well as  logarithmic and exponential \cite{Hendi:2013mka} NLE. It is worth mentioning that, in \cite{Hendi:2010kv} the explicit form of mass of black string, which has a crucial rule in checking the validity of the first law of thermodynamics, is not given. The black string solutions are also explored in the context of ECG \cite{Bueno:2016xff} in \cite{Bakhtiarizadeh:2021vdo,Bakhtiarizadeh:2022bgr} and EQG \cite{Ahmed:2017jod} in \cite{Bakhtiarizadeh:2021hjr}.

The structure of paper is organized as follows. In the next two sections, we will find the black string solution which satisfies the equations of motion of BI/ModMax theory in the context of EG as well as cubic and quartic version of Generalized Quasi-Topological Gravity (GQTG) \cite{Hennigar:2017ego}, and then study the solutions at asymptotic region $ r \to \infty $. In Sec. \ref{near}, we will study the solutions near a black string horizon and obtain, using a Taylor expansion around the horizon, exact expressions for the mass parameter and surface gravity. In Sec. \ref{thermo}, we compute the entropy, ADM mass, angular momentum, total electric charge and the electrostatic potential of solutions and show that the first law of black string thermodynamics holds exactly. Finally, in Sec. \ref{therm} we check the thermal stability of solutins by calculating the specefic heat in canonical ensemble. Section \ref{summ} is also devoted to summary and conclusions. 

\section{Field equations}\label{feq}

Let us start by considering the following action\footnote{Throughout the paper we choose the Newton’s gravitational constant $ G = 1 $.}:
\be
S=\frac{1}{16\pi}\int d^4x \sqrt{-g}(R-2\L-{\cal L}_{\rm GQTG}-{\cal L}_{\rm NLE}),\label{action}
\ee
where $ R $ represents the Ricci scalar, $ \L $ is the negative cosmological constant of AdS space which is related to the AdS radius $ \ell $ as $ \L=-3/{\ell^2} $\footnote{The asymptotically de–Sitter solutions can be obtained by simply taking $ \ell\rightarrow i\ell $ \cite{Awad:2002cz}.}. The Lagrangian of cubic version of GQTG, \ie ECG is given by,
\be
{\cal L}_{\rm ECG}=2 \l {\cal P},
\ee
where the cubic-in-curvature correction to the Einstein-Hilbert action is incorporated in
\bea
{\cal P}=12 R_{a}{}^{c}{}_{b}{}^{d} R_{c}{}^{e}{}_{d}{}^{f} R_{e}{}^{a}{}_{f}{}^{b}+R_{ab}{}^{cd} R_{cd}{}^{ef} R_{ef}{}^{ab}-12 R_{abcd}
R^{ac} R^{bd}+8 R_{a}{}^{b} R_{b}{}^{c} R_{c}{}^{a}.\label{ECGterms}
\eea
The dimensionless ECG coupling constant has been assumed to be non-negative, \ie $ \l \geq 0 $. Einstein gravity is also recovered by setting $ \l = 0 $. Here also the Lagrangian of quartic version of GQTG, \ie EQG reads,
\be
{\cal L}_{\rm EQG}=\sum_{i=1}^{6}{\hat \l}_{(i)} S_{4}^{(i)},
\ee
where $ {\hat \l}_{(i)} $s are the coupling constants of EQG Lagrangian densities. Quartic-in-curvature corrections to the EH action, $ S_{4}^{(i)} $, are called quasi-topological Lagrangian densities, whose analytical expressions read,
\bea
S_{4}^{(1)} &=& - \frac{109}{45} R_{a}{}^{c} R^{ab} R_{b}{}^{d} R_{cd} + \frac{16}{5} R_{ab} R^{ab} R_{cd} R^{cd} + \frac{28}{15} R_{a}{}^{c} R^{ab} R_{bc} R \nonumber \\ 
&& -  \frac{37}{45} R_{ab} R^{ab} \
R^2 + \frac{86}{45} R^{ab} R^{cd} R R_{acbd}  + \frac{1}{18} R^2 R_{abcd} R^{abcd} \nonumber \\ 
&&- \
\frac{208}{15} R_{a}{}^{c} R^{ab} R^{de} \
R_{bdce} + \frac{82}{15} R^{ab} R^{cd} R_{ac}{}^{ef} R_{bdef} -  \frac{7}{30} R R_{ab}{}^{ef} R^{abcd} R_{cdef} \nonumber \\ 
&& -  \frac{8}{9} R_{ab} R^{ab} R_{cdef} \
R^{cdef} -  \frac{1}{3} R^{ab} R_{a}{}^{cde} R_{bc}{}^{fh} R_{defh} + R_{a}{}^{e}{}_{c}{}^{f} R^{abcd} R_{b}{}^{h}{}_{e}{}^{j} R_{dhfj},
\eea
\bea
S_{4}^{(2)} &=& 2 R_{a}{}^{c} R^{ab} R_{b}{}^{d} R_{cd} + \frac{5}{2} R_{ab} R^{ab} R_{cd} \
R^{cd} -  R_{a}{}^{c} R^{ab} R_{bc} R \nonumber \\ 
&& -  \frac{3}{4} R_{ab} R^{ab} R^2 + 4 \
R^{ab} R^{cd} R R_{acbd} + \frac{1}{8} R^2 R_{abcd} R^{abcd} \nonumber \
\\ 
&& - 14 R_{a}{}^{c} R^{ab} R^{de} R_{bdce} + 5 R^{ab} R^{cd} R_{ac}{}^{ef} \
R_{bdef} -  \frac{1}{4} R R_{ab}{}^{ef} \
R^{abcd} R_{cdef} \nonumber \\ 
&& -  \frac{3}{4} R_{ab} R^{ab} R_{cdef} \
R^{cdef} -  R^{ab} R_{a}{}^{cde} R_{bc}{}^{fh} R_{defh} + R_{a}{}^{e}{}_{c}{}^{f} R^{abcd} R_{b}{}^{h}{}_{d}{}^{j} R_{ehfj},
\eea
\bea
S_{4}^{(3)} &=& - \frac{12}{5} R_{a}{}^{c} R^{ab} R_{b}{}^{d} R_{cd} + \frac{2}{5} R_{ab} R^{ab} R_{cd} R^{cd} + \frac{12}{5} R_{a}{}^{c} R^{ab} R_{bc} R \nonumber \\ 
&&-  \frac{1}{5} R_{ab} R^{ab} R^2 + \frac{8}{5} R^{ab} R^{cd} R R_{acbd}  -  \frac{1}{2} R^2 R_{abcd} R^{abcd} \nonumber \\ 
&&- \
\frac{72}{5} R_{a}{}^{c} R^{ab} R^{de} \
R_{bdce} + \frac{48}{5} R^{ab} R^{cd} R_{ac}{}^{ef} R_{bdef}  + \frac{1}{5} R R_{ab}{}^{ef} R^{abcd} \
R_{cdef} \nonumber \\ 
&&+ R_{ab} R^{ab} R_{cdef} \
R^{cdef} - 4 R^{ab} R_{a}{}^{cde} R_{bc}{}^{fh} R_{defh} + R_{ab}{}^{ef} R^{abcd} R_{ce}{}^{hj} \
R_{dfhj},
\eea
\bea
S_{4}^{(4)} &=& - \frac{24}{5} R_{a}{}^{c} R^{ab} R_{b}{}^{d} R_{cd} + \frac{4}{5} R_{ab} R^{ab} R_{cd} R^{cd} + \frac{24}{5} R_{a}{}^{c} R^{ab} R_{bc} R \nonumber \\ 
&&-  \frac{2}{5} R_{ab} R^{ab} R^2 + \frac{16}{5} R^{ab} R^{cd} R \
R_{acbd}  -  R^2 R_{abcd} R^{abcd} \nonumber \\ 
&&-  \frac{144}{5} R_{a}{}^{c} R^{ab} R^{de} R_{bdce} + \frac{96}{5} R^{ab} R^{cd} R_{ac}{}^{ef} R_{bdef} + \frac{2}{5} R R_{ab}{}^{ef} R^{abcd} \
R_{cdef} \nonumber \\ 
&& + 2 R_{ab} R^{ab} R_{cdef} R^{cdef} - 8 R^{ab} R_{a}{}^{cde} \
R_{bc}{}^{fh} R_{defh} + R_{ab}{}^{ef} R^{abcd} R_{cd}{}^{hj} \
R_{efhj},
\eea
\bea
S_{4}^{(5)} &=& - \frac{14}{5} R_{ab} R^{ab} R_{cd} R^{cd} -  \frac{20}{3} R_{a}{}^{b} R_{b}{}^{d} R_{d}{}^{e} R_{e}{}^{a} -  \frac{8}{5} R^{ad} R^{be} R R_{abde} \nonumber \\ 
&&+ \frac{104}{5} R^{ab} R_{c}{}^{e} \
R^{cd} R_{adbe} + R_{ch} R^{ch} R_{abde} R^{abde} + \frac{1}{5} R^2 R_{abde} R^{abde} \nonumber \\ 
&&-  \frac{56}{15} R^{ab} R_{chib} R_{de}{}^{i}{}_{a} R^{dech}  + R_{abd}{}^{c} R^{abde} R_{hife} R^{hif}{}_{c},
\eea
\bea
S_{4}^{(6)} &=& - \frac{308}{15} R_{ab} R^{ab} R_{cd} R^{cd} -  \frac{64}{3} R_{a}{}^{b} R_{b}{}^{d} R_{d}{}^{e} R_{e}{}^{a} + \frac{64}{15} R^{ad} R^{be} R R_{abde} \nonumber \\ 
&&+ \frac{1088}{15} R^{ab} R_{c}{}^{e} \
R^{cd} R_{adbe} + \frac{28}{3} R_{ch} R^{ch} R_{abde} \
R^{abde} -  \frac{8}{15} R^2 R_{abde} R^{abde} \nonumber \\ 
&&-  \frac{224}{15} R^{ab} R_{chib} R_{de}{}^{i}{}_{a} R^{dech} + R_{abcd} R^{abcd} R_{efgh} R^{efgh}.
\eea

In action (\ref{action}), $ {\cal L}_{\rm NLE} $ is also the Lagrangian of corresponding NLE theory: 
\bea\label{NLElag}
{\cal L}_{\rm NLE}=\begin{cases}
	b^2 \lp 1-\sqrt{1+\frac{{\sf S}}{b^2}} \rp, & \text{BI}\\
	2\lp{\sf S}\cosh \g -\sqrt{{\sf S}^2+{\sf P}^2}\sinh \g\rp. & \text{ModMax}
\end{cases}
\eea
Here we consider two classes of NLE theories, namely BI and ModMax. In the Lagrangian of BI theory \cite{Born:1934gh}, $ b $ is a nonlinear parameter. As $ b $ tends to infinity, $ {\cal L} $ reduces to the usual linear Maxwell case $ -{\sf S}/2 $. In the case of ModMax theory \cite{Bandos:2020jsw,Kosyakov:2020wxv}, which is a one-parametric generalization of the Maxwell theory characterized by the dimensionless parameter $ \g $, by setting $ \g = 0 $ one recovers the Maxwell’s case. Also, two electromagnetic invariants are defined as:
\bea\label{elecinvar}
{\sf S}=\frac{1}{2}F_{\m\n}F^{\m\n},\quad{\sf P}=\frac{1}{2}F_{\m\n}{}^\ast\! F^{\m\n}. 
\eea
Here, the field strength $ F_{\m\n} $ is given in terms of the vector potential $ A_{\m} $ by $ F_{\m\n}=2\pa_{[\m} A_{\n]} $, where the Hodge dual of $ F^{\m \n} $ is defined by $ {}^\ast\! F^{\m\n} = (1/2)\e^{\m\n\r\s}F_{\r\s} $.

Let us now present the solution of the EG/ECG/EQG–BI/ModMax theory describing a charged black string in AdS space.  We assume the following metric for spacetime with cylindrical or toroidal horizons \cite{Lemos:1994xp,Lemos:1995cm}, 
\bea\label{sileq}
ds^2=-f(r)g^2(r)\lp \Xi dt -a d\phi \rp ^2+\frac{1}{f(r)}dr^2+\frac{r^2}{\ell^4} \lp a dt -\Xi \ell^2 d\phi \rp ^2+\frac{r^2}{\ell^2}dz^2,\label{met}
\eea
where
\be
\Xi=\sqrt{1+a^2/\ell^2}.
\ee
Here, the constant $ a $ can be regarded as the rotation parameter. The ranges of the time and radial coordinates are $ -\infty<t<\infty, 0\leq r<\infty $.  Hereafter, we suppose that $ 0\leq \phi< 2\pi,-\infty<z<\infty $ which describes a cylindrical horizon with topology $ \mathbb{R}\times S^1 $ and is a stationary black string\footnote{Note that one also can choose the flat torus $ T^2 $ with topology $ S^1\times S^1 $ and the ranges $ 0\leq \phi< 2\pi,0\leq z<2\pi l $, which describes a closed black string. In the case of toroidal horizon, the entropy, mass, angular momentum, and charge of the string are obtained from their respective densities by multiplying them by $ 2\pi l $ \cite{Dehghani:2002rr}.}.

Now, we are in a position to evaluate the field equations of action (\ref{action}) on the ansatz (\ref{met}) and finding the corresponding equations for the function $ f(r) $. The metric is accompanied with the following BI/ModMax field \cite{Lemos:1995cm}:
\bea
&&A_{\m}=h(r)\lp \Xi \d_{\m}^{t}- a \d_{\m}^{\phi} \rp;\nn\\&&F_{tr}=-F_{rt}=-\Xi h'(r),F_{\phi r}=-F_{r\phi}=a h'(r).\label{nonvanF}
\eea
To find the field equations of action (\ref{action}) on the metric (\ref{met}), we use a method introduced in \cite{Hennigar:2016gkm,Bueno:2016lrh} for static and spherically symmetric spacetimes. By considering the action as a functional of $ f $, $ g $ and $ h $, \ie $ S[f,g,h] $, we observe that varying the action with respect to $ f $ yields,
\be
g'(r)=0,
\ee
with the solution $ g = 1 $\footnote{Without loss of generality we can set $ g(r) = 1 $, for simplicity. In general, one can choose $ g = 1/\sqrt{f_{\infty}} $, where $ f_{\infty}=\lim_{r\rightarrow\infty} \frac{\ell^2}{r^2} f(r) $, to normalize the speed of light on the boundary or in the dual CFT to be $ c = 1 $. However, we can set $ g = 1 $ by reparametrization of time ($ t $) and angular ($ \phi $) coordinates of the metric.}. It is also not difficult to show that varying the action with respect to $ h $ yields,
\be
\begin{cases}
	2 h'^3-b^2\lp 2 h'+r h'' \rp =0, & \text{BI}\\
	r h''+2h'=0, & \text{ModMax}
\end{cases}
\ee
with the solution
\bea
h(r)=\begin{cases}\label{hfunction}
	-\frac{Q}{r} \, _2F_1\lp \frac{1}{2},\frac{1}{4};\frac{5}{4};-\frac{Q^2}{b^2 r^4}\rp, & \text{BI}\\
	-\frac{Q}{r}, & \text{ModMax}
\end{cases}
\eea
where $ Q $ is an integration constant, which is related to the electric charge of the black string and $ _2F_1(a,b;c;z) $ is the Gaussian hypergeometric function. Using Eq. (\ref{hfunction}) for $ h(r) $, one can easily find that the Lagrangian of BI/ModMax theory is characterized by the following two invariants:
\bea
{\sf S}=\begin{cases}
	-\frac{Q^2}{r^4} \, _2F_1\lp \frac{1}{2},\frac{1}{4};\frac{5}{4};-\frac{Q^2}{b^2 r^4}\rp, & \text{BI}\\
	-\frac{Q^2}{r^4}, & \text{ModMax}
\end{cases},\qquad {\sf P}=0.
\eea
As a result, the EG, ECG and EQG theories admit solutions characterized by a single function $ f(r) $. The equation $ \d_{g}S = 0 $ yields, after integrating once, the following equations for $ f $ in EG, ECG and EQG theories:
\bea\label{eomEG}
\text{EG:}\;-r f = \frac{1}{3}\L r^3+r_0-n,
\eea
\bea\label{eomECG}
\text{ECG:}\;
	-r f-\l \ls4 f'^3-12 f f' f''-24\frac{f^2 \lp f'-r f''\rp}{r^2}\rs=\frac{1}{3}\L r^3+r_0-n,
\eea
\bea\label{eomEQG}
\text{EQG:}\;
	-r f-\frac{1}{10}K \ls  \frac{3}{r}f'^4+\frac{4}{r^2}ff'^2\lp f'-3r f'' \rp-\frac{24}{r^3} f^2 f' \lp f'-r f''\rp\rs\nn\\=\frac{1}{3}\L r^3+r_0-n,
\eea
where the function $ n(r) $ reads
\bea\label{nequation}
n(r)=
\begin{cases}
\frac{2}{3}\ls b r\lp b r^2-\sqrt{Q^2+b^2r^4} \rp +2\frac{Q^2}{r}\, _2F_1\lp \frac{1}{2},\frac{1}{4};\frac{5}{4};-\frac{Q^2}{b^2 r^4}\rp\rs, & \text{BI}\\
\frac{e^{-\gamma}Q^2}{r}. & \text{ModMax}
\end{cases}
\eea
Here, $ \L=-3/\ell^2 $, $ r_0= M $ is an integration constant (mass parameter) which is related to the mass of black string and a prime denotes a differentiation with respect to $ r $. Imposition of cylindrical symmetry yields a degeneracy among the different theories in (\ref{action}), in that the constant $ K $ is a linear combination of the six EQG coupling constants,
\be\label{EQGcc}
K\equiv -\frac{5}{6}\sum_{i=1}^{6} \l_{(i)},
\ee
where
\bea
\l_{(1)}&=&-\frac{6}{5}{\hat \l}_{(1)},\quad \l_{(2)}=-3{\hat \l}_{(2)},\quad\l_{(3)}=-\frac{12}{5}{\hat \l}_{(3)},\nn\\ \l_{(4)}&=&-\frac{24}{5}{\hat \l}_{(4)},\quad\l_{(5)}=-\frac{24}{5}{\hat \l}_{(5)},\quad\l_{(6)}=-\frac{96}{5}{\hat \l}_{(6)}.
\eea
The combination (\ref{EQGcc}) appears in (\ref{eomEQG}), since each term $ S_4^{(i)} $ has the same contribution to the field equation. This degeneracy shows that there is only one parameter from EQG that can be constrained empirically from analysis of this class of solutions.

\section{Far-region solution}\label{far}

\begin{figure}[htp]
	\begin{picture}(0,0)(0,0)
		\put(115,-9){(a)}
		\put(325,-10){(b)}
	\end{picture}
	\begin{center}
		\includegraphics[width=7.5cm]{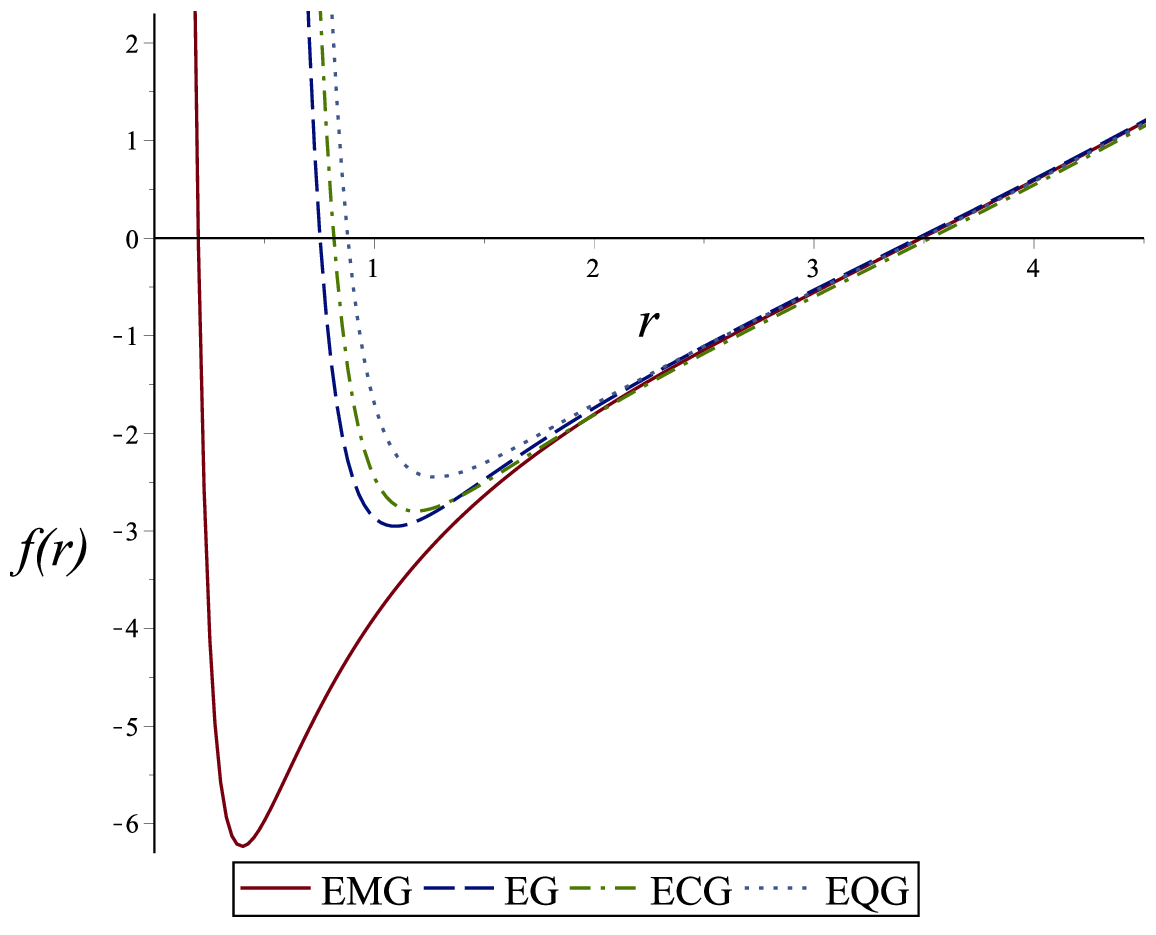} \,\,\,\,\,
		\includegraphics[width=7.5cm]{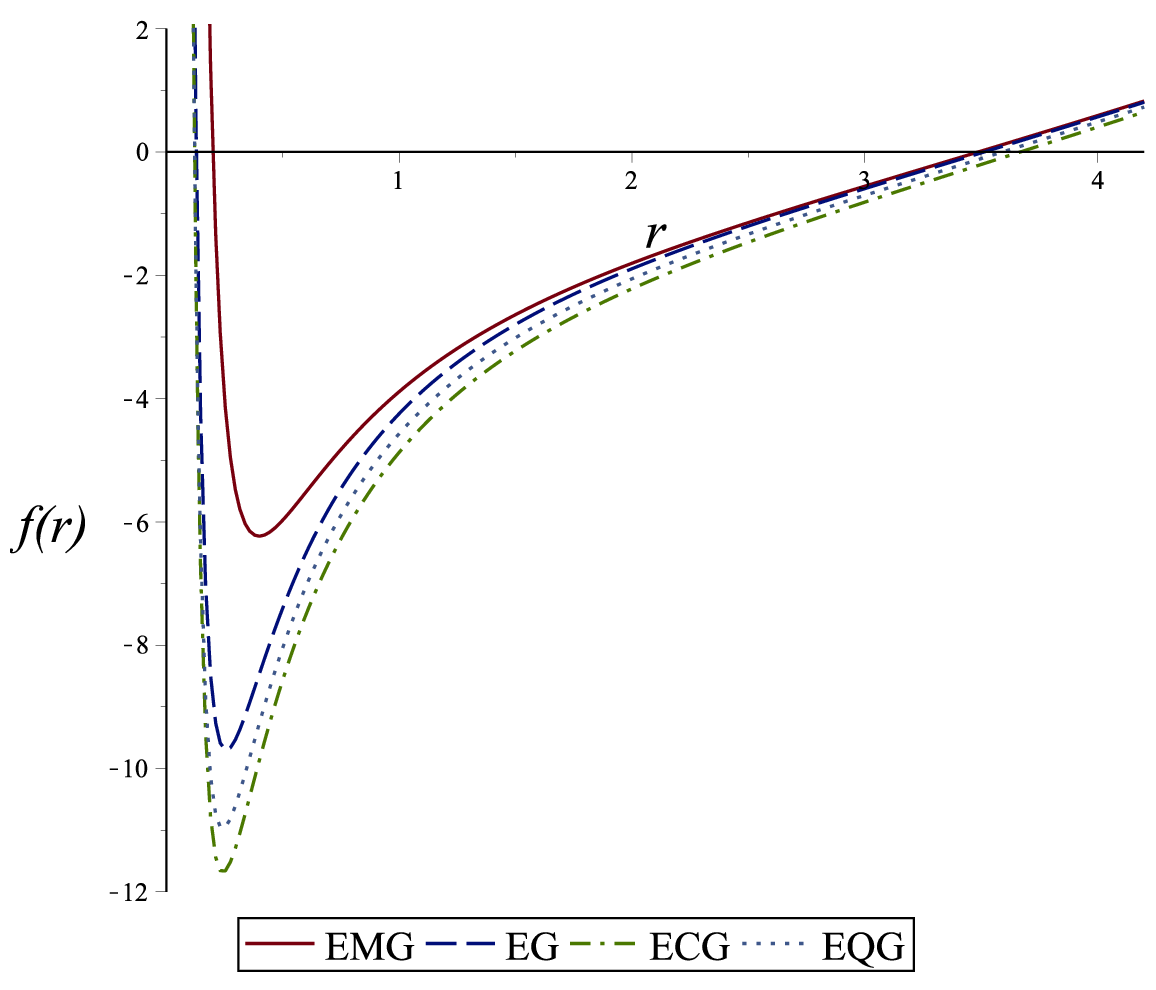} 
		\caption{\small Profile of $ f(r) $ for the EMG and EG/ECG/EQG in (a) BI and (b) ModMax theory. Here, we set the parameters $M=5, \, Q=1, \, \ell=3,\, \l=0.2, K=5, b=0.3$ and $\gamma=2$. }\label{fig1}
	\end{center}
\end{figure}
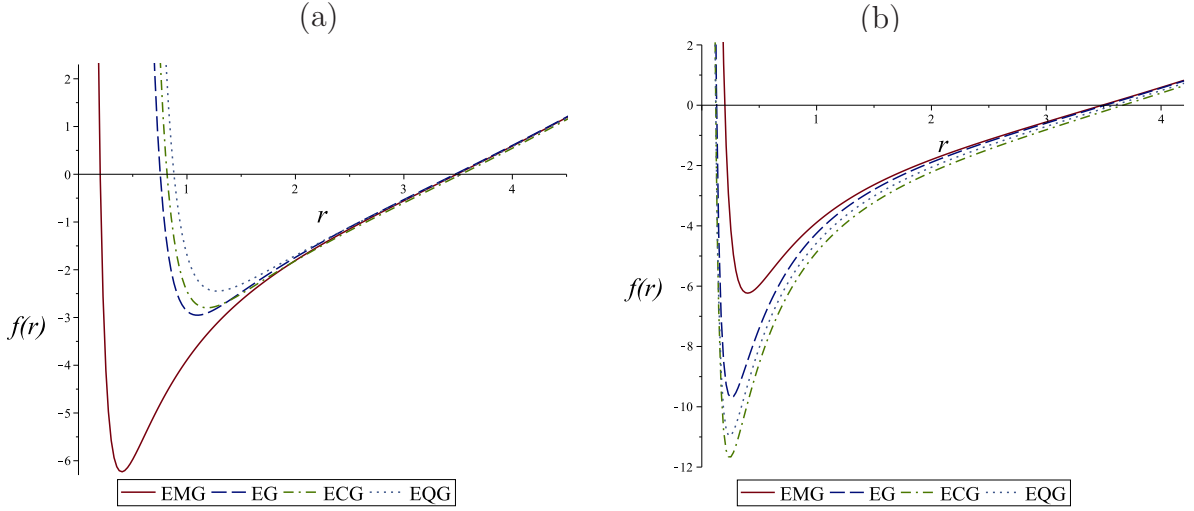

It can be seen from Eq. (\ref{eomEG}) that in the case of EG theory,
\bea\label{f1r}
f(r)=\frac{r^2}{\ell^2}-\frac{M}{r}+\frac{n(r)}{r},
\eea  
where the function $ n(r) $ is given by Eq. (\ref{nequation}), and we have set $ \L=-3/{\ell^2} $ and $ r_0=M $. By turning off the coupling constants of ECG and EQG theory \ie setting $ \l=0 $ in (\ref{eomECG}) and $ K=0 $ in (\ref{eomEQG}), one finds the same result. When $ \l $ and $ K $ are turned on, the asymptotic quantities get corrected in this case. To find these corrections, we examine the large-$ r $ behavior of the solution. In doing so, we consider the metric function $ f $ as a particular solution in the form of a $ 1/r $ expansion, plus the general solution of the corresponding homogeneous equation, \ie
\be\label{serexp}
f=f_{1/r}+f_{h}\quad \text{with} \quad f_{1/r}(r)=\frac{r^2}{\ell_{\rm eff}^2}+\sum_{n=1}^{\infty}\frac{b_{n}}{r^n},
\ee
where $ {\ell_{\rm eff}} $ is the effective radius of AdS space. Substituting the above series expansion into Eqs. (\ref{eomEG})-(\ref{eomEQG}) one finds the following large-$ r $ expansions in EG, ECG and EQG theories,
\be\label{f1rEG}
\text{EG:}\;f_{1/r}(r)=\frac{r^2}{\ell^2}-\frac{M}{r}+\begin{cases}
	\frac{Q^2}{r^2}+\frac{Q^4}{20 b^2 r^6}+\cdots, & \text{BI}\\
	\frac{e^{-\g}Q^2}{r^2}+\cdots, & \text{ModMax}
\end{cases},
\ee
\be\label{f1rECG}
\text{ECG:}\;f_{1/r}(r)=\frac{r^2}{\ell_{\rm eff}^2}-\frac{M}{\lp 1-\frac{48 \l}{\ell_{\rm eff}^4} \rp r}+\begin{cases}
	\frac{Q^2}{\lp 1-\frac{48 \l}{\ell_{\rm eff}^4} \rp r^2}+\frac{Q^4}{20 b^2 \lp 1-\frac{48 \l}{\ell_{\rm eff}^4} \rp r^6}+\cdots, & \text{BI}\\
	\frac{e^{-\g}Q^2}{\lp 1-\frac{48 \l}{\ell_{\rm eff}^4}\rp r^2}+\cdots, & \text{ModMax}
\end{cases},
\ee
\be\label{f1rEQG}
\text{EQG:}\;f_{1/r}(r)=\frac{r^2}{\ell_{\rm eff}^2}-\frac{M}{\lp 1-\frac{32 K}{5 \ell_{\rm eff}^6}\rp r}+\begin{cases}
	\frac{Q^2}{\lp 1-\frac{32 K}{5 \ell_{\rm eff}^6}\rp r^2}+\frac{Q^4}{20 b^2 \lp 1-\frac{32 K}{5 \ell_{\rm eff}^6}\rp r^6}+\cdots, & \text{BI}\\
	\frac{e^{-\g}Q^2}{\lp 1-\frac{32 K}{5 \ell_{\rm eff}^6}\rp r^2}+\cdots, & \text{ModMax}
\end{cases},
\ee
respectively. Here, we have used the fact that $ _2F_1\lp a,b;c;z\rp $ has a convergent series expansion for $ |z| < 1 $ \cite{Dey:2004yt}, and $ \cdots $ denote the terms of order $ 1/r^n $ with $ n>6 $ for BI and $ n>2 $ for ModMax NLE. We can use the above results to plot $ f(r) $ in comparison with Einstein-Maxwell gravity (EMG). Note that, since the horizon is regular, we can in practice continue the solution to the inner region $ f < 0 $. The results are presented in Fig. \ref{fig1}. Note that, these plots are made with the asymptotic expansion of $ f(r) $ and therefore they are not accurate for small $ r $.
By substituting the series expansion (\ref{serexp}) into the field equations (\ref{eomEG})-(\ref{eomEQG}), one also find out that the effective radius of the AdS space, $ {\ell_{\rm eff}} $, is a solution of the following equations
\be\label{leffECG}
\text{ECG:}\;\frac{16 \l}{\ell_{\rm eff}^6}-\frac{1}{\ell_{\rm eff}^2}+\frac{1}{\ell^2}=0,
\ee
\bea\label{leffEQG}
\text{EQG:}\;-\frac{8K}{5 \ell_{\rm eff}^8}-\frac{1}{\ell_{\rm eff}^2}+\frac{1}{\ell^2}=0.
\eea
Here, in the case of EQG theory, we have used the following constraint (by manually converting $ K\to -K $), which have been previously observed in \cite{Bakhtiarizadeh:2021hjr},
\be\label{constr}
-\frac{5 l^6}{128}\leq K \leq 0.
\ee
Note that only this constraint leads to a single root for which the mass parameter and surface gravity will have a smooth limit to the vacuum of Einstein gravity upon sending coupling constant to zero. 

It also has been shown that, for sufficiently large $ r $, the contribution of $ f_h $ in Eq. (\ref{serexp}) is negligible \cite{Bakhtiarizadeh:2021vdo,Bakhtiarizadeh:2021hjr}.

It is easy to show that for the black string solution (\ref{met}) the Ricci scalar and the Kretschmann invariant of the spacetime for EG, ECG, EQG with the both above NLEs are given by 
\bea
R&=&-f''(r)-4 \frac{f'(r)}{r} -2 \frac{f(r)}{r^2},\nn\\
R_{abcd}R^{abcd}&=&{f''}^{2}(r)+\lp\frac{2f'(r)}{r}\rp^2+\lp\frac{2f(r)}{r^2}\rp^2.
\eea
Since other curvature invariants are a function of $ f''(r) $, $ f'(r)/r $ and $ f(r)/r^2 $, it is sufficient to study the Ricci and Kretschmann scalars for investigation of the spacetime curvature. Substituting the metric functions (\ref{f1rEG}), (\ref{f1rECG}) and (\ref{f1rEQG}) one can easily check that for $ r \rightarrow \infty $ the Ricci and Kretschmann scalars go to the values $ R=-12/\ell_{\rm eff}^2=4\L_{\rm eff} $ and $ R_{abcd}R^{abcd}=24/\ell_{\rm eff}^4=(8/3)\L_{\rm eff}^2 $, which by supposing $ {\rm sgn}(\L_{\rm eff}) = {\rm sgn}(\L) $, confirms that our solutions are asymptotically AdS. 

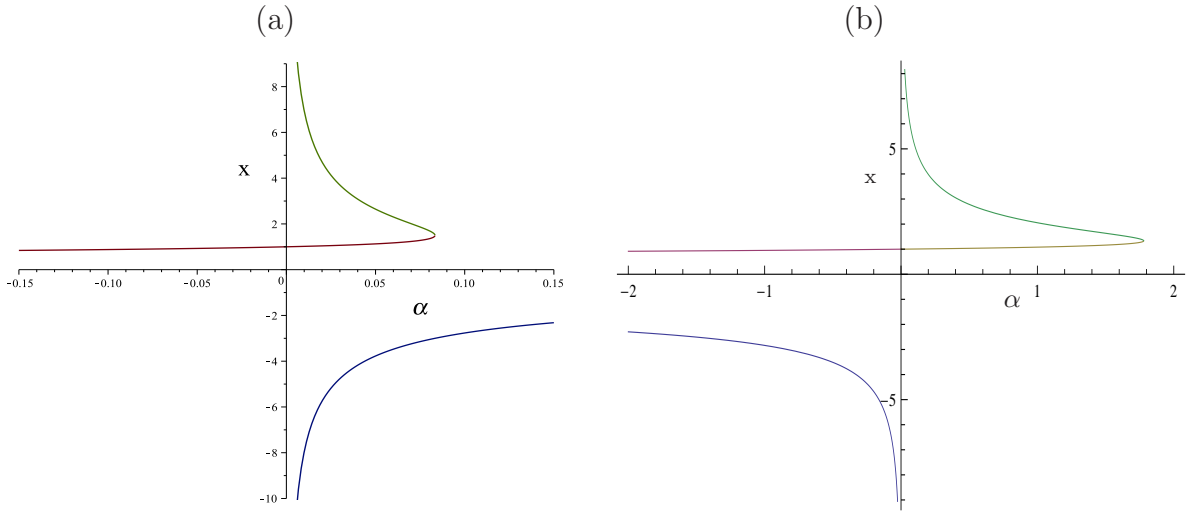
\begin{figure}[htp]
	\begin{picture}(0,0)(0,0)
		\put(100,0){(a)}
		\put(320,0){(b)}
		\put(328,-58){\scriptsize{x}}
		\put(380,-105){\footnotesize{$\alpha$}}
	\end{picture}
	\begin{center}
		\includegraphics[height=6cm,width=7.5cm]{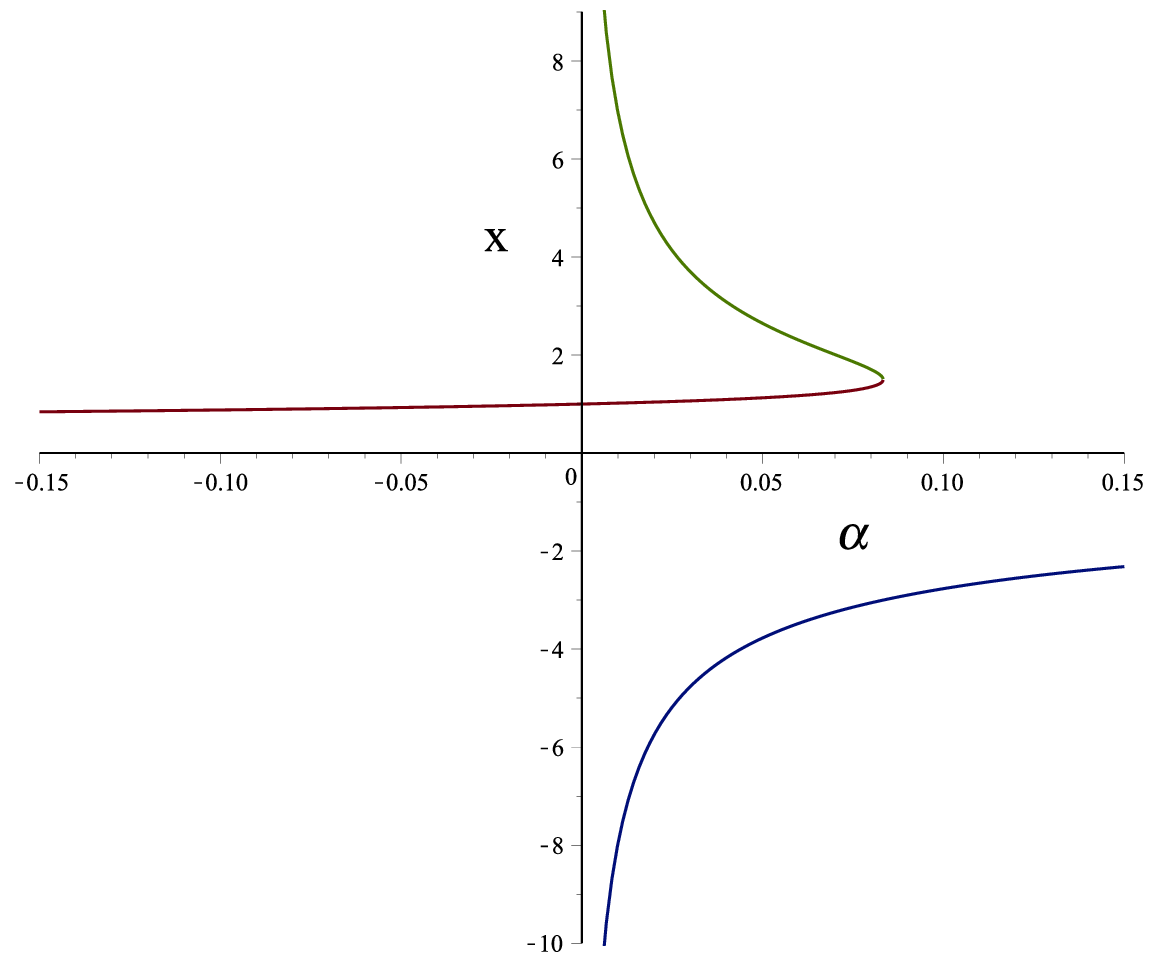} \,\,\,\,\,
		\includegraphics[height=6cm,width=7.5cm]{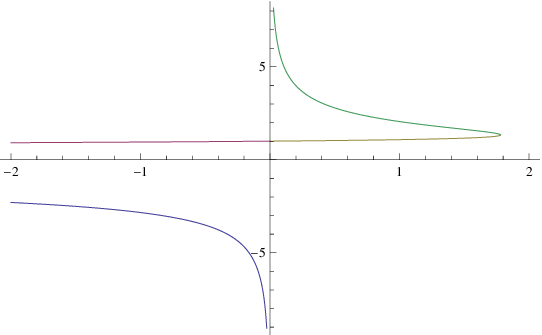} 
		\caption{\small This plot illustrates the solutions for the effective cosmological constant versus  (a) the cubic coupling (b) the quartic coupling where we have made use of the dimensionless parameters $ x = \L_{\rm eff}/\L $ and (a) $ \a = \l \L^2  $ (b) $ \a = K \L^3  $. In both figures we have plotted the $\Delta >0 $ case. The red line in figure (a) indicate the ghost-free branch where  the Einstein theory is recovered in the limit $\lambda \to 0$. In the case of quartic theory in figure (b) there are four branches, however the ghost-free branches are indicated by the yellow and red lines.}\label{fig2}
	\end{center}
\end{figure}

To prove the assumption $ {\rm sgn}(\L_{\rm eff}) = {\rm sgn}(\L) $, let us discuss about the restrictions imposed by Eqs. (\ref{leffECG}) and (\ref{leffEQG}). If we rewrite these equations in terms of cosmological constant, they take the following form:
\be\label{lambdaeffeqECG}
-\frac{16}{27} \l \L_{\rm eff}^3+\frac{\L_{\rm eff}}{3}-\frac{\L}{3}=0,
\ee
\be\label{lambdaeffeqEQG}
-\frac{8}{405} K \L_{\rm eff}^4+\frac{\L_{\rm eff}}{3}-\frac{\L}{3}=0.
\ee
Taking the discriminant of the first equation yields,
\be
\D=-\frac{64\lambda}{729} \left(12 \lambda \Lambda^2- 1\right).
\ee
The discriminant can be either positive, zero, or negative depending on the values of $ \l $ and $ \L $:
\bea
\D&>&0\quad {\rm if} \quad \left(\Lambda <0\land 0<\lambda <\frac{1}{12 \Lambda ^2}\right)\lor (\Lambda =0\land \lambda >0)\nn\\&&~\qquad\lor \left(\Lambda >0\land 0<\lambda <\frac{1}{12 \Lambda ^2}\right),\nn\\
\D&=&0\quad {\rm if} \quad \l=0 \quad {\rm or} \quad \l=\frac{1}{12 \L^2},\nn\\
\D&<&0\quad {\rm if} \quad \left(\Lambda <0\land \left(\lambda <0\lor \lambda >\frac{1}{12 \Lambda ^2}\right)\right)\lor (\Lambda =0\land \lambda <0)\nn\\&&~\qquad\lor \left(\Lambda >0\land \left(\lambda <0\lor \lambda >\frac{1}{12 \Lambda ^2}\right)\right).
\eea
In the case $ \D>0 $, there are three real branches (\ie the theory will have three possible values for $ \L_{\rm eff} $), while there is only one when $ \D<0 $ and there are two branches in the case $ \D=0 $. In general, we find that for $ \Delta>0 $ and any given $ \L $ and $ \l $, there will only be a single branch that is ghost-free which is shown by red line in Fig. \ref{fig2}(a).

Similarly, by taking the discriminant of Eq. (\ref{lambdaeffeqEQG}) we find that,
\be
\D=\frac{64 K^2}{1793613375}\lp -3645+2048 K \L^3\rp.
\ee
Here also the discriminant can be either positive, zero, or negative depending on the values of $ K $ and $ \L $:
\bea
\D&>&0\quad {\rm if} \quad \lp \L <0 \land K <\frac{3645}{2048 \L ^3} \rp  \lor  \lp \L >0 , K >\frac{3645}{2048 \L ^3} \rp,\nn\\
\D&=&0\quad {\rm if} \quad K=0 \quad {\rm or} \quad K=\frac{3645}{2048 \L^3},\nn\\
\D&<&0\quad {\rm if} \quad \lp\L <0 \land \lp K >0 \quad {\rm or}\quad \frac{3645}{2048 \L ^3}<K <0\rp\rp\nn\\&&~\qquad \lor (\L =0\land(K <0\quad {\rm or} \quad K >0))\nn\\&&~\qquad \lor \lp\L >0\land \lp0<K <\frac{3645}{2048 \L ^3}\quad \lor \quad K <0 \rp\rp.
\eea

In the case $ \D>0 $, the theory will have four real branches, in other words four possible values for $ \L_{\rm eff} $, while it will have no real branches when $ \D<0 $ and there are two in the case $ \D=0 $. Here also, we find that for $ \Delta>0 $ and any given $ \L $ and $ K $ there will be two branches that is ghost free. This is shown by yellow and red lines in Fig. \ref{fig2}(b). It is also interesting to note that the ghost-free branch has a smooth limit to the Einstein case as $ \l,K\to 0 $; in other words, the Einstein branch is ghost free.

Furthermore, the ghost-free branch has the property that $ {\rm sgn}(\L_{\rm eff}) = {\rm sgn}(\L) $, meaning, for example, if $ \L < 0 $ then the ghost-free branch will possess AdS asymptotics. 

%Also, in the limit $ r \rightarrow 0 $, both Ricci and Kretschmann scalars diverge, and are finite at $ r\neq 0 $. Therefore, there is a curvature singularity at $ r = 0 $.

\section{Near-horizon solution}\label{near}

The event horizon of black string is a surface which is defined by  $ r = r_h $, at which $ f(r_h) = 0 $ and $ f'(r_h) \geq 0 $. Also, the function must be differentiable at $ r_h $. On the other side, the general definition of surface gravity is
\be
\k_g=\sqrt{-\frac{1}{2}(\nabla^a \chi^b)(\nabla_a \chi_b)}, 
\ee
where the null generator of the black string horizon is given by $ \chi=\pa_t+\Omega\pa_\phi $. Here the angular velocity of the event horizon is given by
\bea 
\Omega=\frac{a}{\ell^2\Xi}. 
\eea
A simple calculation for the spacetime (\ref{met}) gives $ \k_g =f'(r_h)/2\Xi $. 

Plugging a Taylor expansion of the function $ f $ around the horizon (assuming $ f $ to be completely regular there) \ie $ f(r)=\Sigma_{n=1}^{\infty} a_n (r-r_h)^n$ with $ a_n=f^{(n)}(r_h)/n! $, into Eqs. (\ref{eomEG})-(\ref{eomEQG}) and solving them order by order in powers of $ (r-r_h) $, it can be seen that the two lowest-order equations form an algebraic system that is used to fix the mass parameter $ M $ and surface gravity $ \k_g $ in terms of horizon radius $ r_h $, are
\bea\label{MkgEGBI}
\text{EG:}\;\begin{cases}
		\frac{r_h^3}{\ell^2}-M+\frac{2}{3}\ls b r_h\lp b r_h^2-\sqrt{Q^2+b^2r_h^4} \rp +2\frac{Q^2}{r_h}\, _2F_1\lp \frac{1}{2},\frac{1}{4};\frac{5}{4};-\frac{Q^2}{b^2 r_h^4}\rp\rs=0, & \\
		\frac{3 r_h^2}{\ell^2}-2 \k_g \Xi  r_h+2 b\lp b r_h^2-\sqrt{Q^2+b^2r_h^4} \rp=0, & 
	\end{cases}
\eea
\begin{align}\label{MkgECGBI}
\text{ECG:}\;\begin{cases}
	\frac{r_h^3}{\ell^2}-M-32 \l \k_g^3 \Xi^3+\frac{2}{3}\ls b r_h\lp b r_h^2-\sqrt{Q^2+b^2r_h^4} \rp +2\frac{Q^2}{r_h}\, _2F_1\lp \frac{1}{2},\frac{1}{4};\frac{5}{4};-\frac{Q^2}{b^2 r_h^4}\rp\rs=0, & \\
	\frac{3 r_h^2}{\ell^2}-2 \k_g \Xi  r_h+2 b\lp b r_h^2-\sqrt{Q^2+b^2r_h^4} \rp=0, & 
\end{cases}
\end{align}
\begin{align}\label{MkgEQGBI}
\text{EQG:}\;\begin{cases}
	\frac{r_h^3}{\ell^2}-M-\frac{24 K \k_g^4 \Xi^4}{5 r_h}+\frac{2}{3}\ls b r_h\lp b r_h^2-\sqrt{Q^2+b^2r_h^4} \rp +2\frac{Q^2}{r_h}\, _2F_1\lp \frac{1}{2},\frac{1}{4};\frac{5}{4};-\frac{Q^2}{b^2 r_h^4}\rp\rs=0, & \\
	\frac{3 r_h^2}{\ell^2}-2 \k_g \Xi  r_h-\frac{8 K \k_g^4 \Xi^4}{5 r_h^{2}}+2 b\lp b r_h^2-\sqrt{Q^2+b^2r_h^4} \rp=0, & 
\end{cases}
\end{align}
for EG/ECG/EQG-BI theory, and
\bea\label{MkgEGMM} 
	\text{EG:}\;\begin{cases}
		\frac{r_h^3}{\ell^2}-M+\frac{e^{-\g}Q^2}{r_h}=0, & \\
		\frac{3r_h^2}{\ell^2}-2 \k_g \Xi r_h-\frac{e^{-\g}Q^2}{r_h^2}=0, & 
	\end{cases} 
\eea
\bea\label{MkgECGMM} 
\text{ECG:}\;\begin{cases}
	\frac{r_h^3}{\ell^2}-M-32 \l \k_g^3 \Xi^3+\frac{e^{-\g}Q^2}{r_h}=0, & \\
	\frac{3r_h^2}{\ell^2}-2 \k_g \Xi r_h-\frac{e^{-\g}Q^2}{r_h^2}=0, & 
\end{cases} 
\eea
\bea\label{MkgEQGMM}
\text{EQG:}\;\begin{cases}
	\frac{r_h^3}{\ell^2}-M-\frac{24 K \k_g^4 \Xi^4}{5 r_h}+\frac{e^{-\g}Q^2}{r_h}=0, & \\
	\frac{3r_h^2}{\ell^2}-2 \k_g \Xi r_h-\frac{8 K \k_g^4 \Xi^4}{5 r_h^{2}}-\frac{e^{-\g}Q^2}{r_h^2}=0, & 
\end{cases} 
\eea
for EG/ECG/EQG-ModMax theory. Solving the above equations, we get the following quantities for the surface gravity and mass parameter of black string in the case of EG and ECG theories:
\be\label{EGsurgra}
\text{EG:}\; \k_g=\frac{1}{2 \Xi  r_h}\begin{cases}
	\ls r_h^2 \left(2 b^2 +\frac{3}{\ell^2}\right)-2 b \sqrt{Q^2+b^2 r_h^4} \rs, & \text{BI}\\
	\lp\frac{3r_h^2}{\ell^2}-\frac{e^{-\g}Q^2}{r_h^2}\rp, & \text{ModMax}
\end{cases}
\ee
\be\label{EGmass}
\text{EG:}\; M=\begin{cases}
	\frac{r_h^3}{\ell^2}+\frac{2}{3}\ls b r_h\lp b r_h^2-\sqrt{Q^2+b^2r_h^4} \rp +2\frac{Q^2}{r_h}\, _2F_1\lp \frac{1}{2},\frac{1}{4};\frac{5}{4};-\frac{Q^2}{b^2 r_h^4}\rp\rs, & \text{BI}\\
	\frac{r_h^3}{\ell^2}+\frac{e^{-\g}Q^2}{r_h}, & \text{ModMax}
\end{cases}
\ee
\be\label{ECGsurgra}
\text{ECG:}\; \k_g=\frac{1}{2 \Xi  r_h}\begin{cases}
	\ls r_h^2 \left(2 b^2 +\frac{3}{\ell^2}\right)-2 b \sqrt{Q^2+b^2 r_h^4} \rs, & \text{BI}\\
	\lp\frac{3r_h^2}{\ell^2} -\frac{e^{-\g}Q^2}{r_h^2} \rp, & \text{ModMax}
\end{cases}
\ee
\be\label{ECGmass}
\text{ECG:}\; M=\begin{cases}
	\frac{r_h^3}{\ell^2}+\frac{2}{3}\ls b r_h\lp b r_h^2-\sqrt{Q^2+b^2r_h^4} \rp +2\frac{Q^2}{r_h}\, _2F_1\lp \frac{1}{2},\frac{1}{4};\frac{5}{4};-\frac{Q^2}{b^2 r_h^4}\rp\rs\\-\frac{4 \lambda}{r_h^3}\ls r_h^2 \lp2 b^2 +\frac{3}{l^2}\rp-2 b \sqrt{Q^2+b^2 r_h^4}\rs^3, & \text{BI}\\
	\frac{r_h^3}{\ell^2}+\frac{e^{-\g}Q^2}{r_h}-\frac{4 \l }{r_h^{9}} \lp\frac{3r_h^4}{\ell^2} -e^{-\g}Q^2 \rp^3. & \text{ModMax}
\end{cases}
\ee

It can be seen that in the case of ECG theory, for the large values of $ r_h $, the mass parameter becomes  negative, unless $ \l<\ell^4/108 $. This put a new constraint on the upper bound of ECG coupling constant for the black string solutions.

By converting $ K\rightarrow-K $, due to the constraint (\ref{constr}), and solving both equations in (\ref{MkgEQGBI}) and (\ref{MkgEQGMM}) one can obtain the surface gravity and mass parameter of black string, in terms of horizon radius $ r_h $ for EQG theory as
\be\label{EQGsurgra}
\text{EQG:}\; \k_g=\begin{cases}
	\frac{1}{4 \sqrt{3} \ell \Xi K^{5/12}\g'^{1/4}}\lp \frac{\sqrt{\ell}K^{1/4}\g'^{3/4}}{\sqrt{r_h}\a'^{1/6}}-\frac{3^{1/6}\sqrt{\e'}}{\b'^{1/6}} \rp, & \text{BI}\\
	\frac{5^{1/6}}{2 \; 2^{5/6} (3K)^{1/3} \Xi  \sqrt{\ell} \n '^{1/6}}\left(\sqrt{\frac{6 \sqrt{10\n '} r_h^3 \ell^{3/2}}{\sqrt{\th '}}- \th '}+ \sqrt{\th '}\right), & \text{ModMax}
\end{cases}
\ee
\be\label{EQGmass}
\text{EQG:}\; M=\begin{cases}
	\frac{1}{480 r_h}\ls \frac{1}{\ell^4 K^{2/3}\g'}\lp\frac{\sqrt{\ell}K^{1/4}\g'^{3/4}}{\sqrt{r_h}\a'^{1/6}}-\frac{3^{1/6}\sqrt{\e'}}{\b'^{1/6}}\rp+\frac{160 \k'}{\ell^2} \right. \\ \left. +640 Q^2\, _2F_1\lp \frac{1}{2},\frac{1}{4};\frac{5}{4};-\frac{Q^2}{b^2 r_h^4}\rp\rs, & \text{BI}\\
	\frac{r_h^3}{\ell^2}+\frac{e^{-\g}Q^2}{r_h}+\frac{\left(\sqrt{\frac{6 \sqrt{10\n '} r_h^3 \ell^{3/2}}{\sqrt{\th '}}- \th '}+ \sqrt{\th '}\right)^4}{16\; (30K)^{1/3} r_h \ell^2 \n '^{2/3}},& \text{ModMax}
\end{cases}
\ee
where the new parameters have been defined in Appendix \ref{def}. Note that, for higher-order equations, we can treat $ a_2 $ as a free parameter and find the $ n $th coefficient $ a_{n} $, in terms of $ a_2 $, and get a family of solutions with only one free parameter $ a_2 $. 

\begin{figure}[htp]
	\begin{picture}(0,0)(0,0)
		\put(80,0){(a)}
		\put(290,0){(b)}
	\end{picture}
	\begin{center}
		\includegraphics[width=7.5cm]{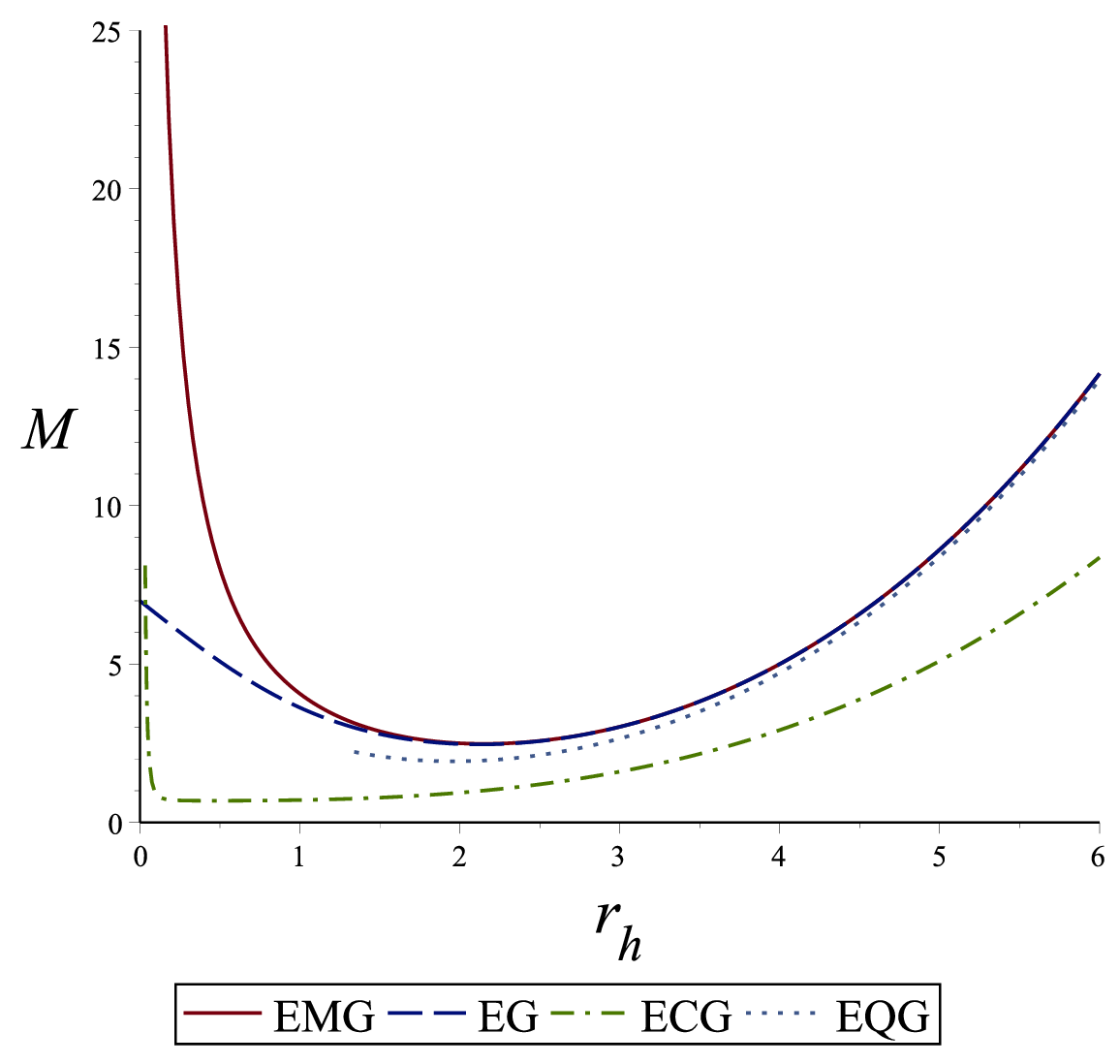}\,\,\,\,
		\includegraphics[width=7.5cm]{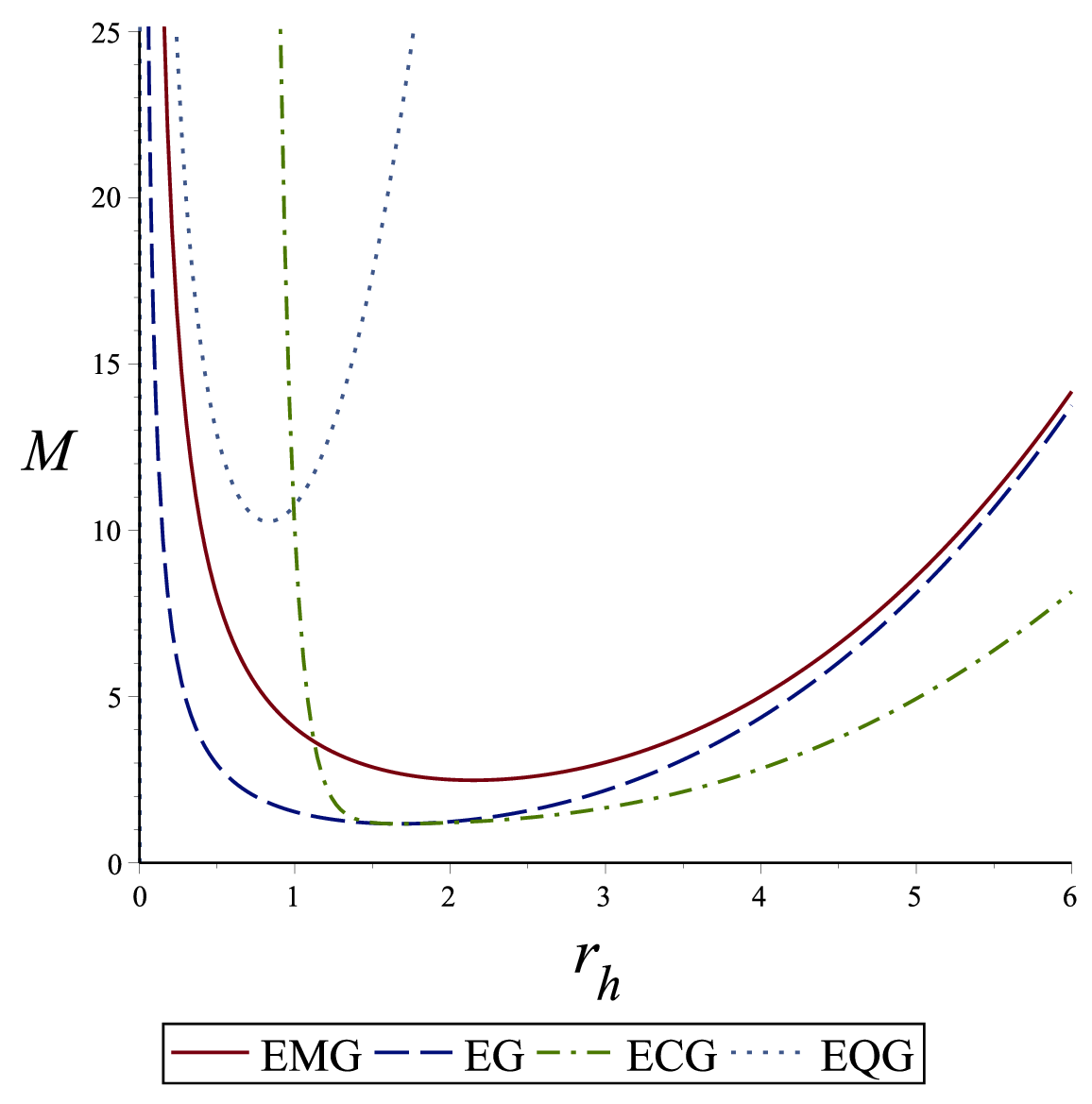}
		 \caption{\small We plot the mass parameter $ M $ as a function of horizon radius $ r_h $ for the EG/ECG/EQG in (a) BI and (b) ModMax theory. Here, we set the parameters $Q=2, \, \ell=4,\, \l=1, K=3,\,\Xi=3/2\sqrt{2}\,(\text{or}\,a=1),\, b=1, $ and $\gamma=1$.}\label{fig3}
	\end{center}
\end{figure}
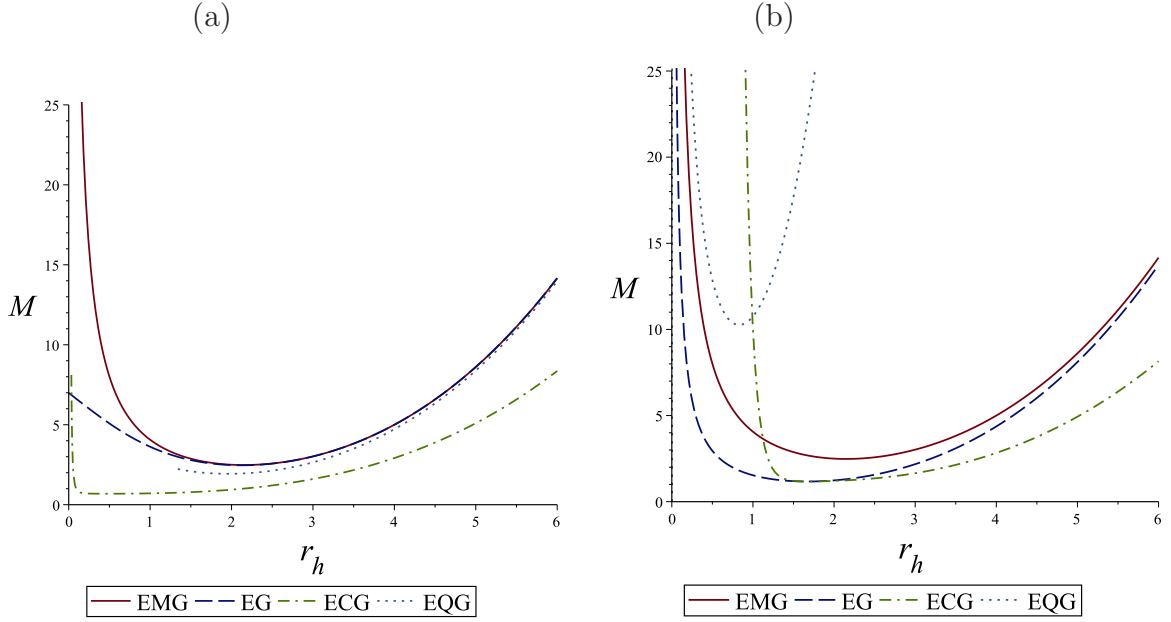

As a cross check, it can be seen that, when $ b \rightarrow \infty $, $ \g \rightarrow 0 $ and $K\to 0$, the above results for the surface gravity and mass parameter, tend to the corresponding ones in RN-AdS solutions of black strings \ie $ \k_g=\lp3r_h^2/\ell^2-Q^2/r_h^2\rp/2 \Xi  r_h $ and $ M=r_h^3/\ell^2+Q^2/r_h $. In Fig. \ref{fig3}, we plot $ M(r_h) $ for BI and ModMax theories, respectively. 

\section{Thermodynamics}\label{thermo}

In this Section, we are going to proceed and identify all the remaining thermodynamic quantities. In the case of EG theory, the black string entropy is given by the Bekenstein’s area law and when higher-curvature corrections are included, \ie in ECG and EQG theories, originates from the Wald entropy formula \cite{Wald:1993nt,Iyer:1994ys}
\be
{\cal S}=\begin{cases}
	\frac{A}{4}, & \text{EG}\\
	-2\pi \int_{H} d^2x\sqrt{\g}\frac{\d {\cal L}}{\d R_{abcd}}\e_{ab}\e_{cd}, & \text{ECG~and~EQG}
\end{cases}
\ee
where $ \frac{\d {\cal L}}{\d R_{abcd}} $ is the Euler-Lagrange derivative of gravitational Lagrangian, $ \g $ is the determinant of the induced metric on the horizon, and $ \e_{ab} $ is the binormal of the horizon, normalized by $ \e_{ab} \e^{ab}=-2 $. This antisymmetric tensor is defined by the following relation:
\be
\triangledown_a \chi_b=\k_g\e_{ab}.
\ee
Using the above equation, the nonzero components of the binormal of horizon of metric (\ref{met}) are given by
\bea
\e_{tr}=-\e_{rt}=-\Xi,\nn\\\e_{r\phi}=-\e_{\phi r}=-a.
\eea
For the metric (\ref{met}) and with a cylindrical horizon placed at $ r = r_h $, one finds the following values for entropy per unit length of black string,
\be
\text{EG:}\;{\cal S}=\frac{\pi r_h^2 \Xi}{2\ell}\qquad\text{for BI and ModMax}
\ee
\be
\text{ECG:}\;{\cal S}=\frac{\pi r_h^2 \Xi}{2\ell}\begin{cases}
	\left\{ 1-\frac{12\l}{r_h^4} \ls r_h^2 \left(2 b^2 +\frac{3}{\ell^2}\right)-2 b \sqrt{Q^2+b^2 r_h^4} \rs^2\right\}, & \text{BI}\\
	\ls 1- \frac{12 \l}{r_h^8} \lp \frac{3 r_h^4}{\ell^2}-e^{-\g}Q^2\rp^2\rs, & \text{ModMax}
\end{cases}
\ee
\be
\text{EQG:}\;{\cal S}=\frac{\pi r_h^2 \Xi}{2\ell}\begin{cases}
	\ls 1+\frac{\lp 3^{2/3} \sqrt{r_h} \alpha'^{1/6} \sqrt{\frac{60 \,3^{1/6} r_h^{7/2} \ell^{5/2} \alpha'^{1/6}}{\sqrt{\gamma'}}-\frac{\delta'}{\beta'^{1/3}}}-\sqrt{3\ell \gamma'} K^{1/4}  \rp^3}{54 \sqrt{r_h\alpha'} K^{1/4} \ell ^3}\rs, & \text{BI}\\
	\ls 1+\frac{\left(\sqrt{\frac{6 \sqrt{10\n '} r_h^3 \ell^{3/2}}{\sqrt{\th '}}- \th '}+ \sqrt{\th '}\right)^3}{3 \ell ^{3/2} \sqrt{10\n '}r_h}\rs, & \text{ModMax}
\end{cases}
\ee 
where the relation $ f'(r_h) =2\Xi \k_g $ and Eqs. (\ref{EGsurgra}), (\ref{ECGsurgra}) and (\ref{EQGsurgra}) have been taken into account to write the final result in terms of horizon radius $ r_h $. The Hawking temperature can easily be written in terms of the horizon radius as  
\bea\label{EGECGtemp}
\text{EG and ECG:}\;T=\frac{1}{4 \pi \Xi  r_h}\begin{cases}
	\ls r_h^2 \left(2 b^2 +\frac{3}{l^2}\right)-2 b  \sqrt{Q^2+b^2 r_h^4} \rs, & \text{BI}\\
	\lp\frac{3r_h^2}{\ell^2}-\frac{e^{-\g}Q^2}{r_h^2}\rp, & \text{ModMax}
\end{cases}
\eea
\bea\label{EQGtemp}
\text{EQG:}\;T=\frac{1}{2 \pi}\begin{cases}
	\frac{1}{4 \sqrt{3}  \ell \Xi K^{5/12}\g'^{1/4}}\lp \frac{\sqrt{\ell}K^{1/4}\g'^{3/4}}{\sqrt{r_h}\a'^{1/6}}-\frac{3^{1/6}\sqrt{\e'}}{\b'^{1/6}}\rp, & \text{BI}\\
	\frac{5^{1/6}}{2 \; 2^{5/6} (3K)^{1/3} \Xi  \sqrt{\ell} \n '^{1/6}}\left(\sqrt{\frac{6 \sqrt{10\n '} r_h^3 \ell^{3/2}}{\sqrt{\th '}}- \th '}+ \sqrt{\th '}\right). & \text{ModMax}
\end{cases}
\eea

For instance, in the case of EG-BI/ModMax, one can easily find the temperature $ T $ as a function of mass parameter $ M $,
\bea\label{Text}
\text{EG:}\;T(M)=\begin{cases}
	\frac{3Mr_h-4Q^2\, _2F_1\lp \frac{1}{2},\frac{1}{4};\frac{5}{4};-\frac{Q^2}{b^2 r_h^4}\rp}{4\pi r_h^3\sqrt{1+\frac{a^2\ls 2b r_h^2 \sqrt{Q^2+b^2 r_h^4}-2b^2r_h^4+3Mr_h-4Q^2\,_2F_1\lp \frac{1}{2},\frac{1}{4};\frac{5}{4};-\frac{Q^2}{b^2 r_h^4}\rp \rs}{3r_h^4}}}, & \text{BI}\\
	\frac{3Mr_h-4Q^2 e^{-\gamma}}{2r_h \sqrt{r_h^4+a^2(Mr_h-Q^2e^{-\g})}}. & \text{ModMax}
\end{cases}
\eea
In the case of ECG/EQG theory, the expressions for $ T(M) $ are so messy to present here. However, it can be seen that, in the case of EG/ECG/EQG theory, there is a critical value of the black string mass parameter,
\bea\label{Mext}
M_{ext}=\begin{cases}
\frac{2 \sqrt{2}  Q^{3/2} \lp 4 b^2 \ell^2 +3 \rp^{1/4} \, _2F_1\lp \frac{1}{2},\frac{1}{4};\frac{5}{4};\frac{-3\lp 4 b^2 \ell^2 +3\rp}{4 b^4 \ell^4}\rp}{3^{3/4}\sqrt{b}\ell}, & \text{BI}\\
\frac{4 Q^{3/2} e^{-3\gamma/4}}{3^{3/4}\sqrt{\ell}}, & \text{ModMax}
	\end{cases}
	\eea
at which, the temperature vanishes, the horizon is degenerate and the black string is extremal. In obtaining Eq. (\ref{Mext}) from (\ref{Text}), we have used the following conditions between the horizon radius $ r_h $ and the physical parameters $ Q, \ell, b, \g $ in the extremal limit of the black string solutions of EG/ECG/EQG theory,
\bea\label{extcond}
\begin{cases}
	r_h^4=\frac{4Q^2b^2\ell^4}{12b^2\ell^2+9}, & \text{BI}\\
	r_h^4=\frac{Q^2e^{-\g}\ell^2}{3}. & \text{ModMax}
\end{cases}
\eea 

We finish this section by calculating the mass and angular momentum of black strings by adding the Gibbons–Hawking boundary term, which removes the divergences of the action (\ref{action}). Here, the suitable boundary action is given by
\be
S_b=\frac{C(\ell_{\rm eff}^2)}{8\pi}\int_{\pa \cal{M}} d^3x \sqrt{-\g} \Th,
\ee
where $ \g $ is the determinant of the induced metric on the boundary and $ \Th $ is the trace of the extrinsic curvature $ \Th_{ab} $ of the boundary. Here, $ C(\ell_{\rm eff}^2) $ is a constant, which depends on the background curvature, and is given by \cite{Bueno:2018xqc,Bueno:2016ypa}
\bea
C(l_{\rm eff}^2)=\begin{cases}
	1, & \text{EG}\\
	1+\frac{48\l}{\ell_{\rm eff}^4}, & \text{ECG}\\
	-\frac{\ell_{\rm eff}^2}{6}{\cal L}\vert_{\rm AdS}, & \text{EQG}\\
\end{cases}
\eea
where $ {\cal L}\vert_{\rm AdS} $ is the Lagrangian of the corresponding theory evaluated on $ {\rm AdS}_4 $ background with curvature scale $ \ell_{\rm eff} $.
We use the counterterm method \cite{Henningson:1998gx,Balasubramanian:1999re} to eliminate the divergences of action. In this approach, we add some local surface integrals to the action to make it finite. The counterterms which makes the action finite are
\be
S_{ct}=\frac{C(\ell_{\rm eff}^2)}{8\pi}\int_{\pa \cal{M}} d^3x \sqrt{-\g} \lp \frac{2}{\ell_{\rm eff}}-\frac{\ell_{\rm eff}}{2} {\cal R}\rp,\label{Sctaction}
\ee
where $ {\cal R} $ is the Ricci scalar for the boundary metric $ \g $. Notice that, $ \ell_{\rm eff} $ is a scale length factor that depends on $ \ell $ and $ \l $ or $ K $, that must reduce to $ \ell $ as $ \l $ or $ K $ go to zero. That is indeed the root of Eqs. (\ref{leffECG}) and (\ref{leffEQG}).
The total action can be written as a linear combination of the action of bulk, boundary, and the counterterm,
\be
S_{total}=S+S_{b}+S_{ct}.
\ee 
Having had the total finite action, one can use the Brown-York definition of stress energy-momentum tensor \cite{Brown:1992br}, by varying the action with respect to boundary metric $ \g_{ab} $, and find a divergence-free stress tensor as
\be\label{enmomtens}
T^{ab}=\frac{1}{8\pi}\lp \Th^{ab}-\Th \g^{ab}+\frac{2}{\ell_{\rm eff}}\g^{ab}-\frac{\ell_{\rm eff}}{2} {\cal G}^{ab}\rp\times\begin{cases}
	1, & \text{EG}\\
	\lp 1+\frac{48\l}{\ell_{\rm eff}^4}\rp, & \text{ECG}\\
	 \lp1-\frac{32 K}{5 \ell_{\rm eff}^6}\rp, & \text{EQG}\\
\end{cases}
\ee
where $ {\cal G}_{ab}={\cal R}_{ab}-{\cal R}\g_{ab}/2 $ is the Einstein tensor of boundary metric $ \g_{ab} $. For asymptotically AdS solutions with flat horizons $ {\cal R}_{abcd}(\g) = 0 $, the only nonvanishing counterterm, is the first term in (\ref{Sctaction}), which yields the stress tensor (\ref{enmomtens}) up to the third term.

Using the above stress tensor, one can define the quasilocal conserved quantities for an asymptotically AdS spacetime. The conserved charge associated to a Killing vector $ \xi_a $ is given by
\be
Q_\xi=\int_{\cal B} d^2x\sqrt{\s} u^a T_{ab}\xi^b,\label{consch}
\ee
where $ u_a =-N\d^{0}_{a} $, while $ N $ and $ \s $ are the lapse function and the metric of a spacelike hypersurface $ {\cal B} $ in the boundary $ \pa {\cal M} $, which appear in the ADM–like decomposition of the boundary metric,
\be
\g_{ij}dx^i dx^j=-N^2 dt^2+\s_{ab}\lp dx^a + V^a dt\rp \lp dx^b + V^b dt \rp.
\ee
Here also $ V^a $ is the shift vector. For the case of black string, the parameters read
\bea\label{lashimet}
N=\sqrt{\frac{r^2 f(r) \left(a^2-\Xi ^2 \ell ^2\right)^2}{\ell ^4 \left(\Xi ^2 r^2-a^2 f(r)\right)}},\quad V^{\phi}=\frac{a \Xi  \left(\ell ^2 f(r)-r^2\right)}{\ell ^2 \left(\Xi ^2 r^2-a^2 f(r)\right)},\nn\\ \s_{ab}dx^a dx^b=\lp\Xi ^2 r^2-a^2 f(r)\rp d\phi^2+\frac{r^2}{\ell^2}dz^2.
\eea
Note that, the explicit form of $ u^{\m} $, which is the unit timelike normal to such a hypersurface is given by
\bea\label{timelike}
u^t = \frac{1}{N} ,u^r = 0, u^\phi=-\frac{V^\phi}{N},u^z=0,
\eea
where $ N $ and $ V^i $ are given in (\ref{lashimet}). 

To obtain the total mass (energy), we should set $ \xi=\pa_t $, \ie the Killing vector conjugate to time coordinate $ t $, and to obtain the angular momentum, we should set $ \xi=\pa_{\ph} $, \ie the Killing vector conjugate to angular coordinate $ \ph $. Using the definition (\ref{consch}) for conserved charges, we find the total mass per unit length of horizon, when $ {\cal B}\to \infty $, to be given by
\be
{\cal M}=\frac{1}{8\ell}\lp 3 \Xi^2-1 \rp M,\label{mas}
\ee
while the angular momentum per unit length of horizon, when $ {\cal B}\to \infty $, is given by
\be
{\cal J}=\frac{3}{8\ell} \Xi a M=\frac{3}{8} \Xi\sqrt{\Xi^2-1} M,\label{ang}
\ee
where in writing the last equality, the Eq. (\ref{sileq}) has been used. Substituting the mass from Eqs. (\ref{EGmass}), (\ref{ECGmass}) and (\ref{EQGmass}) into the above equations, one arrives at the ADM mass and angular momentun in EG, ECG and EQG theories.  

In the following we are going to find the total electric charge per unit length of black string. In the case of electric charge, we first determine the electric field by considering the projections of the electromagnetic field tensor on a spatial hypersurface. We find that
\be
{\cal Q}=\int_{{\cal B}} d^2 x \sqrt{\s} {\cal L}_{\sf S}F^{\m\n}u_{\m}n_{\n},
\ee
where $ {\cal L}_{\sf S}=\pa{\cal L}/2\pa{\sf S} $, the timelike normal unit vector is given by (\ref{timelike}), and the spacelike one is easily found to be,
\bea\label{spacelike}
n^t = 0 ,n^r = \sqrt{f(r)}, n^\phi=0, n^z=0.
\eea
Using the fact that the non-vanishing component of field-strength tensor is $ F^{tr}=E^{r} $, one finds the total electric charge per unit length of black string, when $ {\cal B}\to \infty $, is given as:
\be
{\cal Q}=\begin{cases}
	\frac{\Xi Q}{2\ell}, & \text{BI}\\
	\frac{\Xi e^{-\gamma} Q}{2\ell}. & \text{ModMax}
\end{cases}
\ee

Finally, we calculate the electric potential of the black strings. The electric potential $ \Phi $, measured at infinity with respect to the horizon, is defined by \cite{Dehghani:2002rr}
\be
\Phi=A_a \chi^a\lvert_{r\rightarrow\infty}-A_a\chi^a\lvert_{r=r_h},
\ee
where, as already mentioned, $ \chi $ is the null generator of the horizon. Calculating the above expression yields the following value for electric potential in EG, ECG and EQG theories:
\bea \label{pot}
\Phi=\begin{cases}
	\frac{Q}{\Xi  r_h}\, _2F_1\left(\frac{1}{2},\frac{1}{4};\frac{5}{4};-\frac{Q^2}{b^2 r_h^4 }\right), & \text{BI}\\
	\frac{2 Q}{\Xi  r_h}. & \text{ModMax}
\end{cases}
\eea
We have checked that the above results are in complete agreement with charged black string solution in Maxwell's theory \cite{Lemos:1995cm}. Using the above expressions, one can also show that the first law of thermodynamics for charged rotating black strings, 
\be
d{\cal M}=Td{\cal S}+\Omega d{\cal J}+\Phi d{\cal Q},
\ee
is fully satisfied in the context of EG/ECG/EQG-BI/ModMax theories.

\section{Thermal stability}\label{therm}

\begin{figure}[htp]
	\begin{picture}(0,0)(0,0)
		\put(110,-10){(a)}
		\put(330,-10){(b)}
	\end{picture}
	\includegraphics[height=60mm,width=70mm]{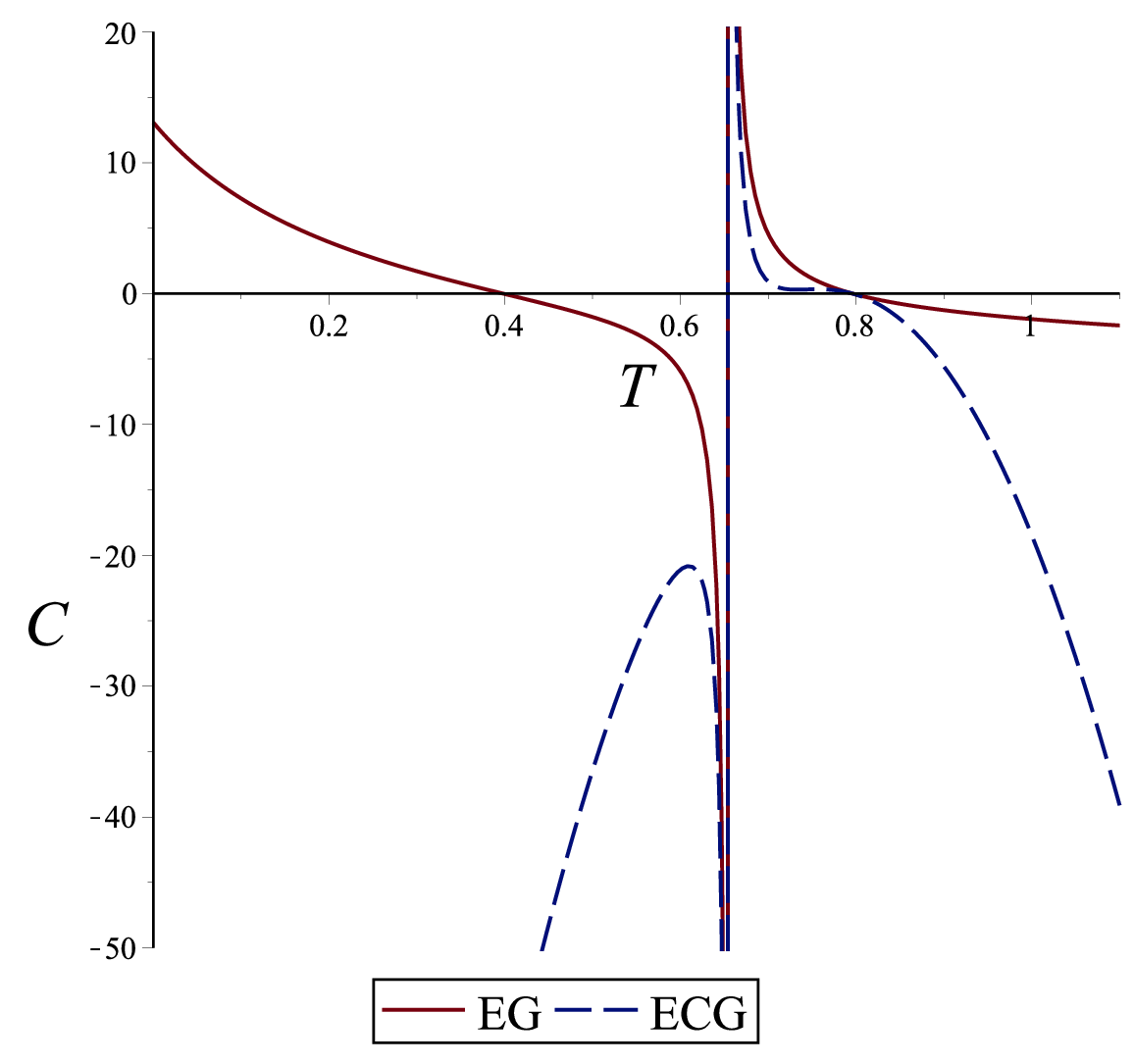} \,
	\includegraphics[height=60mm,width=70mm]{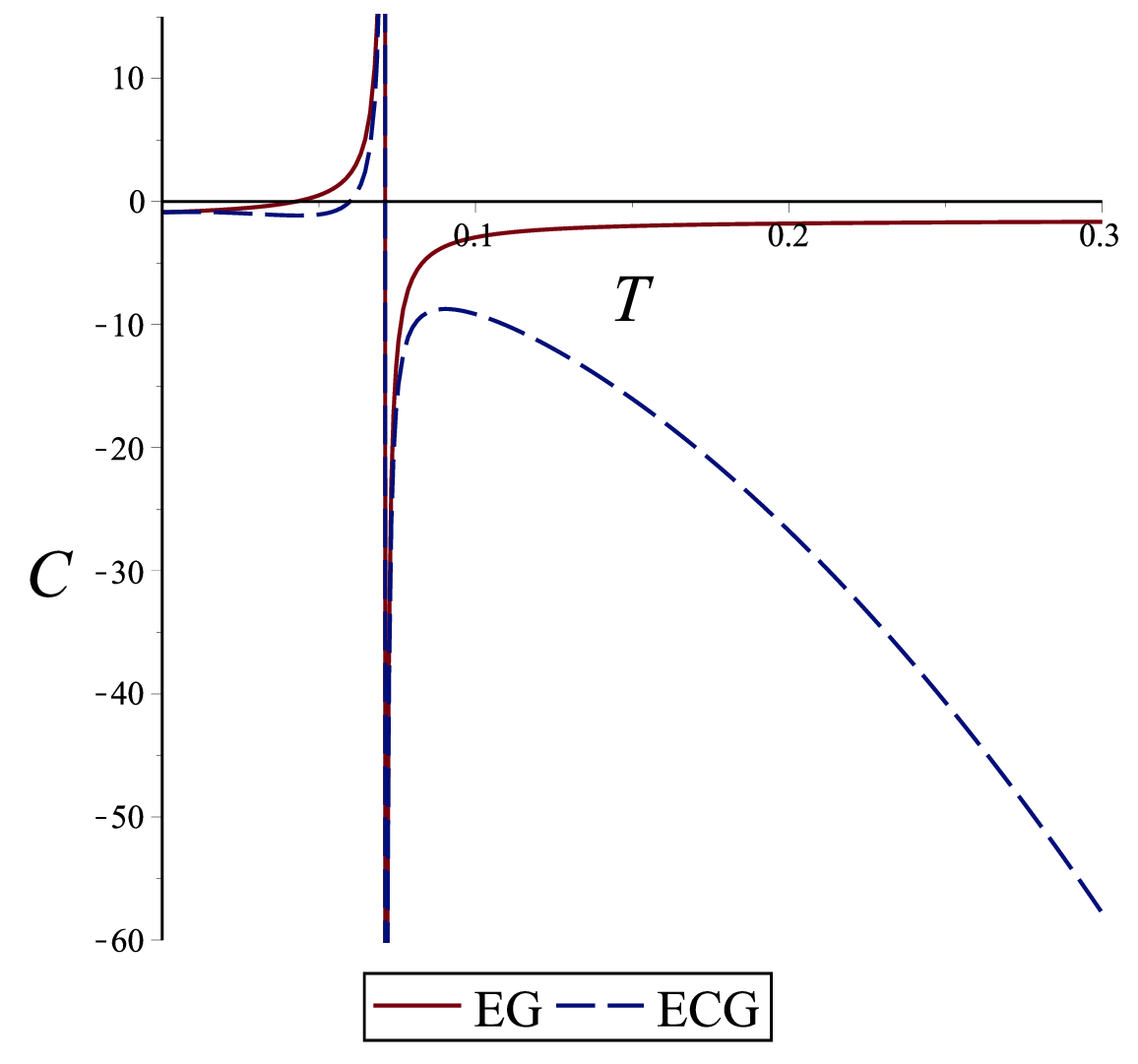}
	\begin{center}
		\caption{\small We plot the specefic heat $ C $ as a function of the temperature $ T $ for the (a) BI and (b) ModMax solution.}\label{fig4}
	\end{center}
\end{figure}
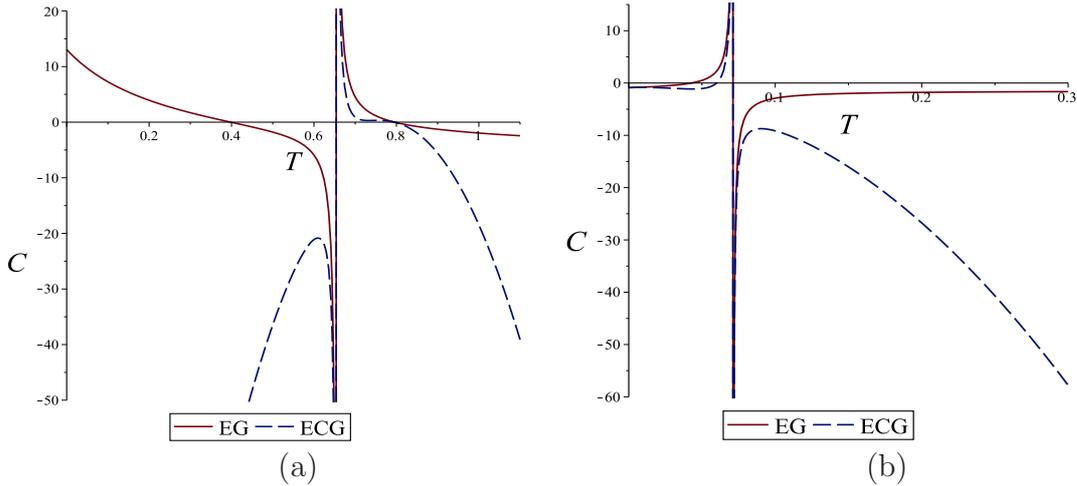

In order to investigate thermal stability of the solution, we find the specific heat, $ C=T\lp\pa {\cal S}/\pa T\rp_{{\cal J},{\cal Q}} $. In canonical ensemble, the angular momentum and charge are the fixed parameters, and therefore
\be
\text{EG:}\;C(T)=\begin{cases}
	\frac{-4\pi \Xi r_h^4 \lp b^2\ell^2+\frac32\rp^2+12\pi^2 \Xi^2 r_h^3 \lp b^2\ell^2+\frac32\rp\ell^2 T-8\pi^3 T^2 \ell^4 \Xi^3 r_h^2 }{8\ell^3\ls \frac12 b^4 \ell^2 r_h^2-\frac{\pi}{2}\lp b^2\ell^2+\frac32\rp \Xi r_h T +\pi^2T^2\Xi^2\ell^2\rs}, & \text{BI}\\
	-\frac{\lp 5\pi T \Xi \ell^2-3 r_h \rp \pi \Xi r_h^2}{6\ell \lp \pi T \Xi \ell^2- r_h \rp},\, & \text{ModMax}
\end{cases}
\ee
\be
\text{ECG:}\;C(T)=\begin{cases}
	\frac{-4\pi \Xi r_h^4 \lp b^2\ell^2+\frac32\rp^2+12\pi^2 \Xi^2 r_h^3 \lp b^2\ell^2+\frac32\rp\ell^2 T-8\pi^3 T^2 \ell^4 \Xi^3 r_h^2 }{8\ell^3\ls \frac12 b^4 \ell^2 r_h^2-\frac{\pi}{2}\lp b^2\ell^2+\frac32\rp \Xi r_h T +\pi^2T^2\Xi^2\ell^2\rs}\\-\frac{48\l \Xi \pi (2\pi T\Xi \ell^2-2b^2\ell^2r_h-3r_h)^2}{\ell^5}, & \text{BI}\\
	-\frac{\lp 5\pi T \Xi \ell^2-3 r_h \rp \pi \Xi r_h^2}{6\ell \lp \pi T \Xi \ell^2- r_h \rp}-\frac{288\pi^3 \l \Xi^3 T^2}{\ell} .\, & \text{ModMax}
\end{cases}
\ee
Since the expresion for EQG is too long, we drop it here. In Fig. \ref{fig4} we plot the specific heat $ C(T) $ as a function of Hawking temperature $ T $ for EG/ECG-BI/ModMax theories. It can be seen that there is a divergency in both cases which means the phase transition for the black string solutions. 

\section{Summary and conclusion}\label{summ}

To summarize, in this paper we have constructed, for the first time, the asymptotically AdS rotating black string solutions in EG, ECG and EQG theories with BI/ModMax source. In order to formulate properly the associated thermodynamics, we have investigated the near-horizon behavior of solutions to find the mass parameter and the Hawking temperature. We also have calculated the entropy, ADM mass, angular momentum, electric charge and the electrostatic potential of solutions and show that the first law of thermodynamics is satisfied for the black string solution in EG/ECG/EQG-BI/ModMax theories.  Also, we have explored the extremality of solutions. Thermal stability of solutions have been investigated which shows a phase transitions.    

It is worth to mention that, although in the case of ModMax theory one finds $ {\sf P}=0 $, the nonlinear characteristic of the black string can be seen from the $ e^{-\gamma} $ terms which appear in the solution. Note that, we do not see the coefficient $ e^{-\gamma} $ as an overall factor. This is more obvious in the case of higher-curvature theories coupled to the ModMax Lagrangian, where in spite of the vanishing $ {\sf P} $, the various factors $ e^{-\gamma} $,  $ e^{-2\gamma} $,  $ e^{-3\gamma} $, $ \cdots $ appear in the solution. 

As a next step, it would be interesting to probe the corresponding solutions in higher dimensions. Finding the solutions in the context of other higher-order gravities is also an interesting feature. 

\appendix

\section{Definition of parameters used in EQG theory}\label{def}

The parameters used in EQG theory are defined as:

\bea
&&\epsilon '\equiv 60 r_h^{7/2} \ell ^{5/2} \sqrt[6]{3\alpha '} \sqrt[3]{\beta '}-\sqrt{\gamma '} \delta ',\nn\\&&\delta '\equiv \sqrt[3]{10} \sqrt[6]{K } \left\{4 \sqrt[3]{30K} r_h^2 \ls r_h^2 \left(2 b^2 \ell ^2+3\right)-2 b \ell ^2 \xi '\rs+\ell ^2 \beta '^{2/3}\right\},\nn\\&&\gamma '\equiv 4\ 30^{2/3} r_h^4 \tau '+\frac{\sqrt[3]{30} r_h^2 \alpha '^{2/3}}{\sqrt[3]{K }}-8\ 30^{2/3} b r_h^2 \ell ^2 \xi ',\nn\\&&\alpha '\equiv 45 r_h^3 \ell ^3+ \sqrt{15\omega '},\nn\\&&\beta '\equiv \sqrt{15\eta '}+45 r_h^6,\nn\\&&\eta '\equiv 135 r_h^{12}-\frac{128 K  \kappa '^3}{\ell ^6},\nn\\&&\omega '\equiv r_h^6 \rho '+256 b K  r_h^4 \ell ^2 \xi ' \psi '-1536 b^2 K  Q^2 r_h^2 \ell ^4 \tau '+1024 b^3 K  Q^2 \ell ^6 \xi ',\nn\\&&\kappa '\equiv r_h^4 \tau '-2 b r_h^2 \ell ^2 \xi ',\nn\\&&\rho '\equiv 135 \ell ^6-128 K  \tau ' \phi ',\nn\\&&\psi '\equiv 4 b^2 \ell ^2 \sigma '+27,\nn\\&&\phi '\equiv 4 b^2 \ell ^2 \zeta '+9,\nn\\&&\sigma '\equiv 4 b^2 \ell ^2+9,\nn\\&&\zeta '\equiv 4 b^2 \ell ^2+3,\nn\\&&\tau '\equiv 2 b^2 \ell ^2+3,\nn\\&&\xi '\equiv \sqrt{Q^2+b^2 r_h^4},\nn\\&&
\m'\equiv 2025 r_h^{12} \ell^6-1920 K \lp 3 r_h^4-e^{-\g}Q^2 \ell ^2 \rp^3,\nn\\&&\n'\equiv \sqrt{\m'}+45 r_h^6 \ell^3,\nn\\&&\varphi'\equiv 4\; 30^{1/3} \lp 3 r_h^4-e^{-\g}Q^2 \ell ^2\rp\nn\\&&
\th'\equiv K^{1/3} \varphi '+\n'^{2/3}.
\eea

%\newpage

\providecommand{\href}[2]{#2}\begingroup\raggedright
\endgroup
\end{document}